\documentclass[aps,prd,twocolumn,showpacs,preprintnumbers,nofootinbib,amsmath,amssymb]{revtex4-2}

\usepackage{graphicx}
\usepackage{dcolumn}
\usepackage{bm}
\usepackage{amsmath}
\usepackage{braket}
\usepackage[hidelinks]{hyperref}
\usepackage{epstopdf}
\usepackage{placeins}
\usepackage{multirow}
\usepackage{makecell}
\usepackage{cleveref}
\usepackage{xcolor}

\newcommand{\beq}{\begin{eqnarray}}
\newcommand{\eeq}{\end{eqnarray}}
\newcommand{\beqnn}{\begin{eqnarray*}}
\newcommand{\eeqnn}{\end{eqnarray*}}
\newcommand{\Tr}{\ensuremath{\mathrm{Tr}}}

\newcommand{\SU}{\mathrm{SU}}
\newcommand{\cool}{\mathrm{cool}}
\newcommand{\sphal}{\mathrm{Sphal}}
\newcommand{\Cov}{\mathrm{Cov}}

\usepackage[font=small, labelfont=bf, textfont={small,it}]{caption}
\def\spose#1{\hbox to 0pt{#1\hss}}
\def\ltapprox{\mathrel{\spose{\lower 3pt\hbox{$\mathchar"218$}}
	\raise 2.0pt\hbox{$\mathchar"13C$}}}

\begin{document}

\title{Sphaleron rate from a modified Backus--Gilbert inversion method}

\author{Claudio Bonanno}
\email{claudio.bonanno@csic.es}
\affiliation{Instituto de F\'isica Te\'orica UAM-CSIC, c/ Nicol\'as Cabrera 13-15, Universidad Aut\'onoma de Madrid, Cantoblanco, E-28049 Madrid, Spain}

\author{Francesco D'Angelo}
\email{francesco.dangelo@phd.unipi.it}
\author{Massimo D'Elia}
\email{massimo.delia@unipi.it}
\author{Lorenzo Maio}
\email{lorenzo.maio@phd.unipi.it}
\author{Manuel Naviglio}
\email{manuel.naviglio@phd.unipi.it}
\affiliation{Dipartimento di Fisica dell'Universit\`a di Pisa \& \\ INFN Sezione di Pisa, Largo Pontecorvo 3, I-56127 Pisa, Italy}

\date{\today}

\begin{abstract}
We compute the sphaleron rate in quenched QCD for a temperature $T \simeq 1.24~T_c$ from the inversion of the Euclidean lattice time correlator of the topological charge density. We explore and compare two different strategies: one follows a new approach proposed in this study and consists in extracting the rate from finite lattice spacing correlators, and then in taking the continuum limit at fixed smoothing radius followed by a zero-smoothing extrapolation; the other follows the traditional approach of extracting the rate after performing such double extrapolation directly on the correlator. In both cases the rate is obtained from a recently-proposed modification of the standard Backus--Gilbert procedure. The two strategies lead to compatible estimates within errors, which are then compared to previous results in the literature at the same or similar temperatures; the new strategy permits to obtain improved results, in terms of 
statistical and systematic uncertainties.
\end{abstract}


\maketitle

\section{Introduction}\label{sec:intro}

The study of real-time topological transitions in finite temperature QCD, the so-called \emph{sphaleron} transitions, has recently attracted much attention from the theoretical community due to its connection to several intriguing phenomenological aspects of the Standard Model, and beyond.

In particular, an extremely interesting role is played by the \emph{sphaleron rate}
\beq\label{eq:rate_def}
\begin{aligned}
\Gamma_\sphal &= \underset{t_{\mathrm{M}}\to\infty}{\underset{V_s\to\infty}{\lim}} \, \frac{1}{V_s t_{\mathrm{M}}}\left\langle\left[\int_0^{t_{\mathrm{M}}} d t_{\mathrm{M}}' \int_{V_s} d^3x \, q(t_{\mathrm{M}}', \vec{x})\right]^2\right\rangle\\
\\[-1em]
\\[-1em]
\\[-1em]
&=\int d t_{\mathrm{M}} d^3x \braket{q(t_{\mathrm{M}},\vec{x}) q(0,\vec{0})},
\end{aligned}
\eeq
where $t_{\mathrm{M}}$ is the real Minkowski time and
\beq\label{eq:topchargedens_cont}
q(x) = \frac{1}{32 \pi^2} \varepsilon_{\mu\nu\rho\sigma}\Tr\{ G^{\mu\nu}(x) G^{\rho\sigma}(x)\}
\eeq
is the QCD topological charge density, expressed in terms of the gluon field strength $G_{\mu\nu} \equiv \partial_\mu A_\nu - \partial_\nu A_\mu +i [A_\mu, A_\nu]$.

For example, a non-vanishing sphaleron rate drives local fluctuations in the difference between the left and right axial quark numbers $N_L - N_R$, being $q(x)$ coupled to the divergence of the axial quark current $J_5^\mu = \bar{\psi} \gamma^\mu\gamma_5 \psi$ due to the anomalous breaking of $\mathrm{U}(1)_{\mathrm{A}}$. When imbalances in the axial quark number due to sphaleron transitions are created in the presence of strong background magnetic fields, such as those generated for short times during heavy-ion collisions, they lead to the so-called \emph{Chiral Magnetic Effect}~\cite{Fukushima:2008xe, Kharzeev:2013ffa, Astrakhantsev:2019zkr, Almirante:2023wmt}, which is one of the most intriguing predictions for the quark-gluon plasma. Another example of the importance of $\Gamma_\sphal$ comes instead from Beyond Standard Model phenomenology. Indeed, the sphaleron rate has been recently recognized as an essential input for the computation of the rate of thermal axion production in the early Universe via axion-pion scattering~\cite{Notari:2022zxo}.

Because of such prominent phenomenological role, the computation of the QCD sphaleron rate at finite temperature has been tackled in recent years in the literature, although so far just restricting to the quenched case~\cite{Kotov:2018aaa, Kotov:2019bt, Altenkort:2020axj, BarrosoMancha:2022mbj} (i.e., the quarkless pure $\SU(3)$ gauge theory). Due to the non-perturbative nature of sphaleron dynamics, being driven by topological excitations, numerical Monte Carlo (MC) simulations on the lattice are a natural tool to compute $\Gamma_{\sphal}$. Being the latter based on the Euclidean formulation of QCD, the real-time definition of $\Gamma_\sphal$ in Eq.~\eqref{eq:rate_def} cannot be directly used to compute this quantity numerically. However, using the Kubo formula, one can express the rate in terms of the slope of the \emph{spectral density} $\rho(\omega)$ in the zero-frequency limit (here $T$ is the temperature):
\beq\label{eq:kubo}
\Gamma_\sphal = 2T \lim_{\omega \to 0} \frac{\rho(\omega)}{\omega}.
\eeq
The quantity $\rho(\omega)$ is related to the Euclidean topological charge density time-correlator,
\beq\label{eq:def_tcorr}
G(t) \equiv \int d^3 x \braket{q(t, \vec{x}) q(0, \vec{0})},
\eeq
with $t$ the imaginary Euclidean time, via the following integral relation~\cite{Meyer:2011gj}:
\beq\label{eq:rho_def}
G(t) = - \int_0^{\infty} \frac{d\omega}{\pi} \rho(\omega) \frac{\cosh\left[\dfrac{\omega}{2T} - \omega t\right]}{\sinh\left[\dfrac{\omega}{2T}\right]}.
\eeq
It is clear that, to extract $\Gamma_\sphal$ from lattice simulations, the main difficulty is constituted by the inversion of Eq.~\eqref{eq:rho_def} to obtain $\rho(\omega)$ from $G(t)$.

Inverse problems are a general class of problems which arise in several different intriguing physical contexts (see, e.g., Refs.~\cite{Rothkopf:2022fyo,Aarts:2023vsf} for recent reviews on the topic), and are well known to be ill-posed (or at least ill-conditioned). Despite these mathematical difficulties, in the literature several different strategies have been devised to find approximate solutions to inverse problems, such as methods based on sum rules~\cite{Boito:2022njs}, on Bayesian approaches~\cite{Horak:2021syv,DelDebbio:2021whr,Candido:2023nnb}, on perturbative-motivated ans\"{a}tze of the spectral density~\cite{Altenkort:2020axj,Altenkort:2020fgs,Altenkort:2022yhb,Altenkort:2023oms}, on the Tikhonov regularization~\cite{Tikhonov:1963aaa,Astrakhantsev:2018oue,Astrakhantsev:2019zkr}, or on the model-independent Backus--Gilbert approach~\cite{BackusGilbert1968:aaa,Brandt:2015aqk,Brandt:2015sxa,Hansen:2019idp,ExtendedTwistedMassCollaborationETMC:2022sta,Frezzotti:2023nun,Evangelista:2023fmt}, which is the one we will adopt in this work.

In general terms, the Backus--Gilbert method allows to numerically reconstruct the spectral density $\rho(\omega)$ in terms of a linear combination of the values of the correlator $G(t)$ determined on the lattice, whose coefficients are obtained from the minimization of a suitable functional. The core of the method, thus, relies on the specific strategy pursued to fix such coefficients. Here, we will rely on the one recently introduced in Ref.~\cite{Hansen:2019idp}, which is a modification of the original proposal of Ref.~\cite{BackusGilbert1968:aaa}. Although considering also other approaches to solve the inverse problem in Eq.~\eqref{eq:rho_def} goes beyond the scopes of this paper, we stress here that this method is expected to yield equivalent results with respect to other proposals. As a matter of fact, the Backus--Gilbert approach of~\cite{Hansen:2019idp} has been shown to be equivalent, within the framework of Bayesian approaches, to a Gaussian Process~\cite{10.1093/gji/ggz520} (see the extended discussion in Ref.~\cite{ExtendedTwistedMassCollaborationETMC:2022sta} on this point). Also the sum-rule-based method of Ref.~\cite{Boito:2022njs} builds on ideas originally developed in~\cite{Hansen:2019idp}. Finally, the original Backus--Gilbert method~\cite{BackusGilbert1968:aaa} has been shown to agree with the Tikhonov regularization~\cite{Astrakhantsev:2018oue,Astrakhantsev:2019zkr}, while in Ref.~\cite{Altenkort:2022yhb} the Backus--Gilbert approach of Refs.~\cite{Brandt:2015aqk,Brandt:2015sxa} 
has been shown to give results consistent with those obtained by a different strategy, based on the fit of lattice data to perturbative-inspired models for the spectral density.

Another aspect that has to be treated with some care is the lattice determination of the topological charge density correlator. As a matter of fact, due to UV noise, it is customary to determine topological quantities from smoothened configurations obtained from the application of some smoothing algorithm. After smoothing, UV fluctuations are suppressed up to a scale known as the \emph{smoothing radius}, which is proportional to the square root of the amount of smoothing performed. However, since smoothing modifies short-distance fluctuations, computing $G(t)$ using Eq.~\eqref{eq:def_tcorr} from determinations of $q(x)$ obtained on smoothened gauge fields unavoidably modifies the behavior of the correlator at small times.

A possible strategy to overcome this issue, adopted in Refs.~\cite{Kotov:2018aaa,Altenkort:2020axj}, is to perform a double extrapolation of the correlator: first one performs a continuum extrapolation of the lattice correlator at fixed smoothing radius; finally, one extrapolates continuum determinations of $G(t)$ towards the zero-smoothing-radius limit. The latter approach, however, has the drawback of working only for sufficiently large Euclidean times $t$.

Indeed, the range of smoothing radii that can be considered for the zero-smoothing extrapolation is bounded from below (as a minimum amount of smoothing is necessary to ensure that we are correctly identifying the topological background of the configuration) and from above (as the smoothing radius needs to be smaller than the time distance $t$ between the correlated sources). Therefore, such range closes for smaller values of $t$. While this fact does not constitute a total obstruction for the extraction of $\Gamma_\sphal$ from the Backus--Gilbert method (being it related to the zero-frequency behavior of $\rho(\omega)/\omega$, which is dominated by the behavior of $G(t)$ at larger times), it makes the reconstruction of the spectral density noisier, making it more difficult to obtain reliable results for $\Gamma_\sphal$.

In this work, instead, we propose a different approach, namely, to move the double extrapolation on the rate. In practice, we determine the rate from the correlators obtained at finite lattice spacing and smoothing radius, and then we perform the double extrapolation outlined earlier directly on $\Gamma_\sphal$. The main idea behind this strategy is the expectation that the reconstruction could be more accurate, compared to the one done on the double-extrapolated correlator, i.e., affected by smaller statistical and systematic uncertainties. In principle, this new approach is less theoretically justified, as the perturbative argument discussed in Ref.~\cite{Altenkort:2020fgs} suggests that the integral relation in Eq.~\eqref{eq:rho_def} can be distorted for asymptotically large frequencies when considering a finite-smoothing-radius correlator; however, one can heuristically expect this problem to be less important in the opposite limit $\omega \to 0$, which is the one interesting for the sphaleron rate computation and related to the infrared (IR) behavior of the correlator, which is less affected by smoothing.

In particular, it is reasonable to expect that there is a regime, if smoothing is not excessively prolonged, where the finite UV cut-off introduced by the non-zero smoothing radius does not have a significant impact on the obtained results for $\Gamma_\sphal$, much like what happens, e.g., to the topological susceptibility computed from the gradient flow as a function of the flow time. If it is possible to identify such a regime, one can expect the sphaleron rate to approach a plateau as a function of the smoothing radius, which signals an effective separation between the UV scale of the smoothened fluctuations and the IR scale of the topological fluctuations relevant to $\Gamma_\sphal$. In the following we will show that this is indeed the case.

The goal of our work is to compare the two methods here outlined, in view of an application to the more computationally demanding case of full QCD. Therefore, we focus on one value of the temperature, namely $T \simeq 1.24~T_c \simeq 357$~MeV, and we perform our study in quenched QCD, where our results can also be compared with other independent determinations in the literature.

This paper is organized as follows: in Sec.~\ref{sec:inversion_method} we explain in details our numerical setup, focusing on the computation of the correlator and on the inversion method to extract the rate; in Sec.~\ref{sec:rate_results} we present our numerical results for the rate; in Sec.~\ref{sec:conclu} we draw our conclusions and discuss future perspectives.

\section{Numerical setup}\label{sec:inversion_method}

In this section we will discuss our numerical setup, the parameters of our simulations and the methods employed to compute the topological charge density correlators and to perform their inversion to obtain the sphaleron rate.

\subsection{Lattice action and parameters}\label{sec:lat_action}

We discretize the Euclidean pure-$\SU(3)$ gauge action $S_{\mathrm{YM}}= (1/4g^2)\int d^4x \, \Tr\{G_{\mu\nu}(x)G_{\mu\nu}(x)\} $ on a $N_s^3 \times N_t$ lattice with lattice spacing $a$ using the standard Wilson lattice gauge action
\beq\label{eq:wilson_action} 
S_{\mathrm{W}} = - \frac{\beta}{3} \sum_{n,\mu>\nu}\Re \Tr \left[ \Pi_{\mu\nu}(n) \right],
\eeq
where $\beta = 6/g^2$ is the bare inverse gauge coupling and $\Pi_{\mu\nu}(n) \equiv U_\mu(n) U_\nu(n+\hat{\mu})U^\dagger_\mu(n+\hat{\nu})U^\dagger_\nu(n)$ is the plaquette.

We performed simulations for 4 values of $\beta$, corresponding to 4 values of the lattice spacing $a$, following a Line of Constant Physics (LCP) where the spatial volume $\left[a(\beta) N_s\right]^3 \simeq [1.66(2)~\text{fm}]^3$, the aspect ratio $N_s/N_t=3$ and the temperature $T = \left[a(\beta) N_t\right]^{-1} \simeq 357(5)~\text{MeV} \simeq 1.24(2)~T_c$ were kept fixed for each gauge ensemble. Scale setting was performed according to the lattice spacing determinations in units of the Sommer parameter $r_0$ reported in Ref.~\cite{Necco:2001xg}, and all simulations parameters are summarized in Tab.~\ref{tab:simulation_summary}. We also checked that using the different parameterization of $a(\beta)/r_0$ of Ref.~\cite{Francis:2015lha} gave perfectly agreeing results within the $\sim 1\%$ precision with which the lattice spacing is determined (see App.~\ref{app:scale_setting}).

Configurations were generated adopting a mixture of the standard local Over-Relaxation (OR)~\cite{Creutz:1987xi} and Over-Heat-Bath (HB)~\cite{Creutz:1980zw,Kennedy:1985nu} algorithms, both implemented \emph{\`a l\`a}
Cabibbo--Marinari~\cite{Cabibbo:1982zn}, i.e., updating all the 3
diagonal $\SU(2)$ subgroups of $\SU(3)$. In particular, our single MC updating step consisted of 1 lattice sweep of HB followed by 4 lattice sweeps of OR. The measure of the topological charge density correlator was performed every 20 MC steps, and the total statistics employed to compute $G(t)$ is reported in Tab.~\ref{tab:simulation_summary}.

\begin{table}[!t]
\begin{center}
\begin{tabular}{|c|c|c|c|c|c|c|}
\hline
$N_s$ & $N_t$ & $\beta$ & $a/r_0$ & $L/r_0$ & $r_0T$ & Stat.\\
\hline
36 & 12 & 6.440 & 0.09742(97) & 0.8554(86) & 3.507(35) & 80k \\
42 & 14 & 6.559 & 0.08364(84) & 0.8540(85) & 3.513(35) & 10k \\
48 & 16 & 6.665 & 0.07309(73) & 0.8551(86) & 3.508(35) & 16k \\
60 & 20 & 6.836 & 0.05846(58) & 0.8553(86) & 3.508(35) & 5k  \\
\hline
\end{tabular}
\end{center}
\begin{center}
\begin{tabular}{|c|c|c|c|c|c|}
\hline
$N_t$ & $\beta$ & $a$~[fm] & $L$~[fm] & $T$~[MeV] & $T/T_c$ \\
\hline
12 & 6.440 & 0.04598(67) & 1.655(24) & 357.6(5.2) & 1.244(18) \\
14 & 6.559 & 0.03948(58) & 1.658(24) & 357.0(5.2) & 1.242(18) \\
16 & 6.665 & 0.03450(50) & 1.656(24) & 357.5(5.2) & 1.244(18) \\
20 & 6.836 & 0.02759(40) & 1.656(24) & 357.6(5.2) & 1.244(18) \\
\hline
\end{tabular}
\end{center}
\caption{Summary of simulation parameters. The scale was set with a $\sim 1 \%$ accuracy according to the determination of $a(\beta)/r_0$ in the range $5.7 \le \beta \le 6.92$ reported in Eq.~(2.6) of Ref.~\cite{Necco:2001xg}. In order to convert quantities in units of $r_0$ to physical $\mathrm{MeV}/\,\mathrm{fm}$ units, we used the value of the Sommer parameter $r_0 = 0.472(5)~\mathrm{fm}$ given in Ref.~\cite{Sommer:2014mea}. Finally, to express the temperature in units of the the critical one, we used the latest and to-date most accurate determination $T_c = 287.4(7)~\mathrm{MeV}$~\cite{Borsanyi:2022xml}, converted from $w_0$ into physical units using~\cite{Borsanyi:2012zs}. The total statistics collected is expressed in thousands (k), and measures were collected every 20 MC updating steps (defined in the text).}
\label{tab:simulation_summary}
\end{table}

\subsection{Lattice topological charge density correlator and smoothing}

We discretized the continuum topological charge density in Eq.~\eqref{eq:topchargedens_cont} using the standard \emph{clover} definition, which is the simplest lattice discretization with definite parity:
\beq
q_L(n) = \frac{-1}{2^9 \pi^2}\sum_{\mu\nu\rho\sigma=\pm1}^{\pm4}\varepsilon_{\mu\nu\rho\sigma}
\Tr\left\{\Pi_{\mu\nu}(n)\Pi_{\rho\sigma}(n)\right\},
\eeq
where it is understood that $\varepsilon_{(-\mu)\nu\rho\sigma} = - \varepsilon_{\mu\nu\rho\sigma}$.

To obtain the correlator in dimensionless physical units, we measured the time profile $Q_L(n_t)$ of the lattice topological charge $Q_L$
\beq\label{eq:topcharge_profile}
Q_L(n_t) = \sum_{\vec{n}} q_L(n_t,\vec{n}), \qquad Q_L = \sum_{n_t} Q_L(n_t),
\eeq
and computed
\beq\label{eq:topchargedens_lat}
\frac{G_L(tT)}{T^5} = \frac{N_t^5}{N_s^3}\braket{Q_L(n_{t,1})Q_L(n_{t,2})},
\eeq
where the physical time separation between the sources is given by
\begin{equation}
\begin{gathered}
tT = 
\begin{cases}
\vert n_{t,1} - n_{t,2} \vert/ N_t   , & \vert n_{t,1} - n_{t,2} \vert \le N_t/2,\\
\\[-1em]
\\[-1em]
1- \vert n_{t,1} - n_{t,2} \vert/ N_t, & \vert n_{t,1} - n_{t,2} \vert   > N_t/2.
\end{cases}
\end{gathered}
\end{equation}
Note that it is sufficient to compute the correlator up to $tT=0.5$, as $G_L(tT)=G_L(1-tT)$.

The topological charge profiles entering Eq.~\eqref{eq:topchargedens_lat} are computed after smoothing, in order to ensure that we consider only correlations of fluctuations 
of physical origin. Indeed, the lattice topological charge $Q_L$ in Eq.~\eqref{eq:topcharge_profile} renormalizes multiplicatively as follows~\cite{Campostrini:1988cy,Vicari:2008jw}:
\beq
Q_L = Z_Q(\beta) Q,
\eeq
where $Q$ is the continuum integer-valued topological charge. Moreover, the two-point function of the lattice topological charge density 
contains short-distance UV artefacts, leading for instance to the appearance of additive 
renormalizations in higher-order cumulants of the topological charge distribution~\cite{DiVecchia:1981aev,DElia:2003zne}, which become dominant in the continuum limit, overcoming the physical signal. Being such effects related to fluctuations on the scale of the UV cut-off, which are dumped by smoothing, computing the lattice topological charge density correlator on smoothened configurations removes such renormalizations, ensuring that one is correctly considering only correlations of physical relevance.

Several smoothing algorithms have been adopted in the literature, such as cooling~\cite{Berg:1981nw,Iwasaki:1983bv,Itoh:1984pr,Teper:1985rb,Ilgenfritz:1985dz,Campostrini:1989dh,Alles:2000sc}, stout smearing~\cite{APE:1987ehd, Morningstar:2003gk} or gradient flow~\cite{Luscher:2009eq, Luscher:2010iy}. All choices give consistent results when properly matched to each other~\cite{Alles:2000sc, Bonati:2014tqa, Alexandrou:2015yba}.

In this work we choose cooling for its simplicity and numerical cheapness. One cooling step consists in a sweep of the lattice where we align each link $U_\mu(n)$ to its local staple. Iterating the cooling steps drives the Wilson action~\eqref{eq:wilson_action} closer to a local minimum, thus dumping UV fluctuations while leaving the global topological content of the field configuration unaltered.

We recall that, while in the continuum $G(t) < 0$ for every $t>0$ because of reflection positivity~\cite{Alles:1997ae,Vicari:1999xx,Horvath:2005cv,Vicari:2008jw,Chowdhury:2012sq,Fukaya:2015ara,Mazur:2020hvt}, on the lattice this property is violated for smaller time separations, because the sources entering in the lattice correlator are smoothed. As a matter of fact, the lattice correlator $G_L$ is negative only when the time separation between the sources is larger that the smoothing radius; otherwise, it will be positive. Of course, after the double extrapolation (i.e., continuum limit followed by zero-smoothing limit), the negativity of the correlator is recovered.

\subsection{Inversion Method}\label{sec:backus-gilbert}

Once the correlation function $G(t)$ is computed, Eq.~\eqref{eq:rho_def} has to be inverted to extract the spectral function $\rho(\omega)$ and then compute the sphaleron rate using Eq.~\eqref{eq:kubo}. Let us rewrite Eq.~\eqref{eq:rho_def} as:
\beq\label{eq:rho_with_Kprime}
G(t) = - \int_0^{\infty} \frac{d\omega}{\pi} \frac{\rho(\omega)}{f(\omega) } K_t^\prime(\omega),
\eeq
where $f(\omega)$ is an arbitrary function, and where we redefined the basis function as
\beq\label{eq:def_Kp}
K'_t(\omega) \equiv  f(\omega) \frac{\cosh[\omega/(2T) - \omega t]}{\sinh[\omega/(2T)]}.
\eeq
In the case of Backus--Gilbert techniques, one constructs the estimator $\bar{\rho}(\omega)$ of the spectral function as:
\beq\label{eq:Estimator}
\bar{\rho}(\bar{\omega})= - \pi f(\bar{\omega}) \sum_{t=0}^{1/T} g_t(\bar{\omega}) G(t),
\eeq
where $g_t$ are unknown coefficients to be determined. The advantage of this formulation is that we can set $f(\omega) = \omega$ and $\bar{\omega} = 0$, so that we are able to directly estimate from the correlator the ratio $\rho(\omega)/\omega$ in the limit $\omega\rightarrow 0$:
\beq
\left[\frac{\bar{\rho}(\bar{\omega})}{\bar{\omega}}\right]_{\bar{\omega}\,=\,0}= - \pi \sum_{t=0}^{1/T} g_t(0) G(t).
\eeq
This is, apart from an overall factor, the sphaleron rate according to the Kubo formula~\eqref{eq:kubo}.

Combining Eqs.~\eqref{eq:rho_with_Kprime} and~\eqref{eq:Estimator}, one obtains the following relation between the estimator $\bar{\rho}(\bar{\omega})$ and the physical spectral function $\rho(\omega)$:
\beq\label{eq:rel_smeared}
\frac{\bar{\rho}(\bar{\omega})}{\bar{\omega}} = \int_0^{\infty} d\omega \Delta(\omega,\bar{\omega}) \frac{\rho(\omega)}{\omega},
\eeq
where 
\beq\label{eq:resolution}
\Delta(\omega,\bar{\omega})= \sum_{t=0}^{1/T} g_t(\bar{\omega}) K'_t(\omega)
\eeq
is the so-called \emph{resolution function}.

From Eq.~\eqref{eq:rel_smeared} it follows that, assuming a resolution function normalized to 1, if $\Delta(\omega, \bar{\omega})$ has a sharp peak around $\bar{\omega}$ as a function of $\omega$, then $\bar{\rho}$ is a good approximation of the actual spectral function $\rho$. This is particularly evident in the limit in which $\Delta(\omega, \bar{\omega})$ tends to a Dirac delta-function $\delta(\omega-\bar{\omega})$: in this case the relation $\bar{\rho}(\bar{\omega}) = \rho(\bar{\omega})$ holds exactly. Clearly, in a real calculation the resolution function will have a peak of finite width around $\bar{\omega}$. Thus, the estimator $\bar{\rho}(\bar{\omega})$ will actually be an average of the spectral function over such a region around $\bar{\omega}$. This means that the larger the width of the resolution function is, the less faithfully we are able to reconstruct the actual spectral density $\rho$ from $\bar{\rho}$. It is therefore clear that the strategy used to fix the shape of the resolution function in terms of the unknown $g_t$ coefficients plays a crucial role in determining the quality of our estimation of the spectral density via $\bar{\rho}$.

To compute the coefficients $g_t$, we apply the modified Backus--Gilbert regularization method recently proposed in~\cite{Hansen:2019idp}. This approach consists in minimizing a functional depending on the difference between the resolution function $\Delta(\omega,\bar{\omega})$ and some chosen target function $\delta(\omega,\bar{\omega})$, whose shape is fixed on the basis of physical considerations. Since such procedure is typically extremely noisy, it is customary to regularize it by adding to the minimized functional a term related to the statistical error on the reconstructed quantity.

In our case, the functional $F[g_t]$ that is minimized to determine $g_t$ takes the following form:
\beq\label{eq:backus_gilbert_functional_tot}
F[g_t] = (1-\lambda) A_\alpha[g_t] +  \frac{\lambda}{\mathcal{C}}B[g_t], \,\quad  \lambda\in [0,1),
\eeq
where $\mathcal{C}$ is a normalization factor proportional to the square of the value of the correlator in a fixed point (here we used $\mathcal{C}=G(tT=0.5)^2$), $\lambda$ is a free parameter whose role will be discussed later, and $A_\alpha$ and $B$ are suitable functionals depending on $g_t$.

The functional $A_\alpha$ is related to the distance between the resolution and the given target function $\delta(\omega, \bar{\omega})$:
\beq\label{eq:backus_gilbert_functional_A}
A_\alpha[g_t] = \int_0^{\infty} d\omega \, [\Delta(\omega,\bar{\omega}) - \delta(\omega,\bar{\omega})]^2 \,e^{\alpha \omega}, \,\, \alpha <2.
\eeq
As proposed in~\cite{ExtendedTwistedMassCollaborationETMC:2022sta}, the square distance between $\Delta(\omega,\bar{\omega})$ and $\delta(\omega, \bar{\omega})$ is further multiplied by an exponentially growing factor to promote larger frequencies in the integral defining $A_\alpha[g_t]$. This is justified by the known one-loop perturbative result for $\rho(\omega)$, which predicts that $\rho(\omega)$ diverges as a power-law in $\omega$ at large frequencies~\cite{Laine:2011xm}. In our analysis we used $\alpha=2^-$, i.e., $\alpha= 1.99$.

The second functional is proportional to the uncertainty on the final quantity (i.e., the spectral density):
\beq\label{eq:backus_gilbert_functional_B}
B[g_t] = \sum_{t,t'=0}^{1/T} \Cov_{t,t'} \, g_t g_{t'},
\eeq
where $\Cov_{t,t'} = \braket{[G(t)-\braket{G(t)}][G(t')-\braket{G(t')}]}$ denotes the covariance matrix of the correlator.

As proposed in~\cite{Almirante:2023wmt}, we used the pseudo-Gaussian target function
\beq\label{eq:target}
\delta_\sigma(\omega, \bar{\omega}=0) = \left(\frac{2}{\sigma \pi}\right)^2 \frac{\omega}{\sinh(\omega/\sigma)},
\eeq
which depends on the free parameter $\sigma$, the \emph{smearing width}, related to the width of the target function. The choice of $\sigma$ directly reflects on the width of the resolution function obtained after the minimization procedure outlined above, and thus on the quality of our estimation of the spectral function. Choosing larger values of $\sigma$ will yield smaller errors on the rate, as coefficients $g_t$ will have smaller fluctuations, but the results will also be less physically reliable. On the other hand, the more peaked the target function is chosen, the noisier our determination of the rate will be. In our analysis, we chose $\sigma/T = 1.75$, but we also checked that choosing other values gave compatible results for the rate within the errors, so that any systematics related to the choice of the smearing width is well under control (more details on this point can be found in App.~\ref{app:rate_vs_sigma}). Therefore, we fixed $\sigma/T=1.75$ for all analyzed ensembles, meaning that we used such value both for the correlators we obtained at finite lattice spacing and for the one obtained in the continuum limit. For this value of the width of the target function, the observed relative deviation at the peak between the resolution and the target function was at most of $\sim 5\%$ for $\lambda=0$.

Once $\rho(\bar{\omega})/\bar{\omega}\vert_{\bar{\omega}=0}$ is obtained from the Backus-Gilbert inversion method, we compute the sphaleron rate using Eq.~\eqref{eq:kubo}. We do so for several values of the free parameter $\lambda \in [0,1)$ appearing in the functional~\eqref{eq:backus_gilbert_functional_tot}. When $\lambda\to0$, i.e., when we neglect the regulator term $B[q_t]$, statistical errors on the sphaleron rate explode, since the inversion problem defining $\rho$ is ill-posed, and coefficients $g_t$ will have sizeable fluctuations. As $\lambda$ is increased, the inversion problem gets regularized and errors on $\Gamma_\sphal$ decrease. However, when $\lambda\to 1$, we are neglecting the contribution of the functional $A_\alpha$, and the resulting resolution function we get from our minimization procedure is practically unconstrained, and can vary sizeably even upon a small variation of $\lambda$. Therefore, in this regime, the result of our inversion cannot be trusted from a physical point of view, and will be dominated by systematic effects. Therefore, to provide a correct estimation of the sphaleron rate, we chose $\lambda$ in order to stay within the statistically-dominated region, and we included any observed systematic variation of the rate within this region in our final error budget.

More precisely, this is the procedure we have followed to estimate our final error on the rate. First, we compute the sphaleron rate as a function of the quantity
\beq\label{eq:def_d}
d[g_t](\lambda) \equiv \sqrt{\frac{A_0[g_t]}{B[g_t]}},
\eeq
where the statistical error on $\Gamma_\sphal$ was computed, for each value of $d[g_t](\lambda)$, from a bootstrap analysis carried over $O(1000)$ bootstrap resamplings. According to our previous discussion, it is clear that, when $d[g_t](\lambda)$ is small, we are reasonably within the statistically-dominated regime.

Then, we select a point in the statistically-dominated region, corresponding to a value $d[g_t](\lambda_1)\ll 1$, whose central value will be the central value of our final estimate of $\Gamma_\sphal$, and whose statistical error will be the statistical error on our determination of the rate.

Finally, we select a second point deeper in the statistically dominated regime $d[g_t](\lambda_2) < d[g_t](\lambda_1)$ to estimate possible systematics. More precisely, we compute a systematic error which is proportional to the difference between the central values of the rates obtained for $\lambda_1$ and $\lambda_2$ (according to Eqs.~(37) and~(38) of~\cite{ExtendedTwistedMassCollaborationETMC:2022sta}). In the end, the final error on $\Gamma_\sphal(\lambda_1)$ is obtained summing in quadrature the systematic and the statistical errors.

\section{Numerical results for the sphaleron rate}\label{sec:rate_results}

In this section we will show and discuss our results for the sphaleron rate, obtained by using two different strategies:
the standard one, based on the inversion of the double-extrapolated time correlator of the topological charge density; and the 
new one, proposed in this paper, which consists of performing the double extrapolation directly on the sphaleron rate itself, obtained from the inversion of finite-lattice-spacing and finite-smoothing-radius correlators. In both cases, we make use of the the modified Backus--Gilbert method described in Sec.~\ref{sec:backus-gilbert}.

\subsection{Rate from the double-extrapolated correlator}\label{sec:double_extr_corr}

Let us start by discussing our result for the sphaleron rate obtained from the inversion of the double-extrapolated correlator.

The first step is, of course, to extrapolate the lattice correlator $G_L(tT)/T^5$ towards the continuum limit at fixed smoothing radius. To do so, with our setup it is sufficient to keep $n_\cool/N_t^2$ fixed for each lattice spacing. As a matter of fact, the relation between the smoothing radius $r_s$ in lattice units and the number of cooling steps $n_\cool$ is given by~\cite{Bonati:2014tqa}:
\beq\label{eq:smooth_rad_cool}
\frac{r_s}{a} \simeq \sqrt{\frac{8 n_\cool}{3}} \,.
\eeq
Therefore, $n_\cool/N_t^2 \propto (r_sT)^2$. Since $n_\cool$ can only assume integer values, in order to keep $n_\cool/N_t^2$ fixed for each ensemble we performed a spline cubic interpolation of our correlators at non-integer values of $n_\cool$.

Moreover, in order to compute the continuum limit of $G(tT)$, we also need the same physical time separation $tT$ for each lattice spacing. Therefore, for each value of $n_\cool$, we also interpolated the correlators obtained on coarser lattices to the values of $tT$ obtainable on the finest one. Also in this case, we did a spline cubic interpolation of the correlators, similarly to what has been done in Ref.~\cite{Altenkort:2020axj}.

\begin{figure}[!htb]
\centering
\includegraphics[scale=0.45]{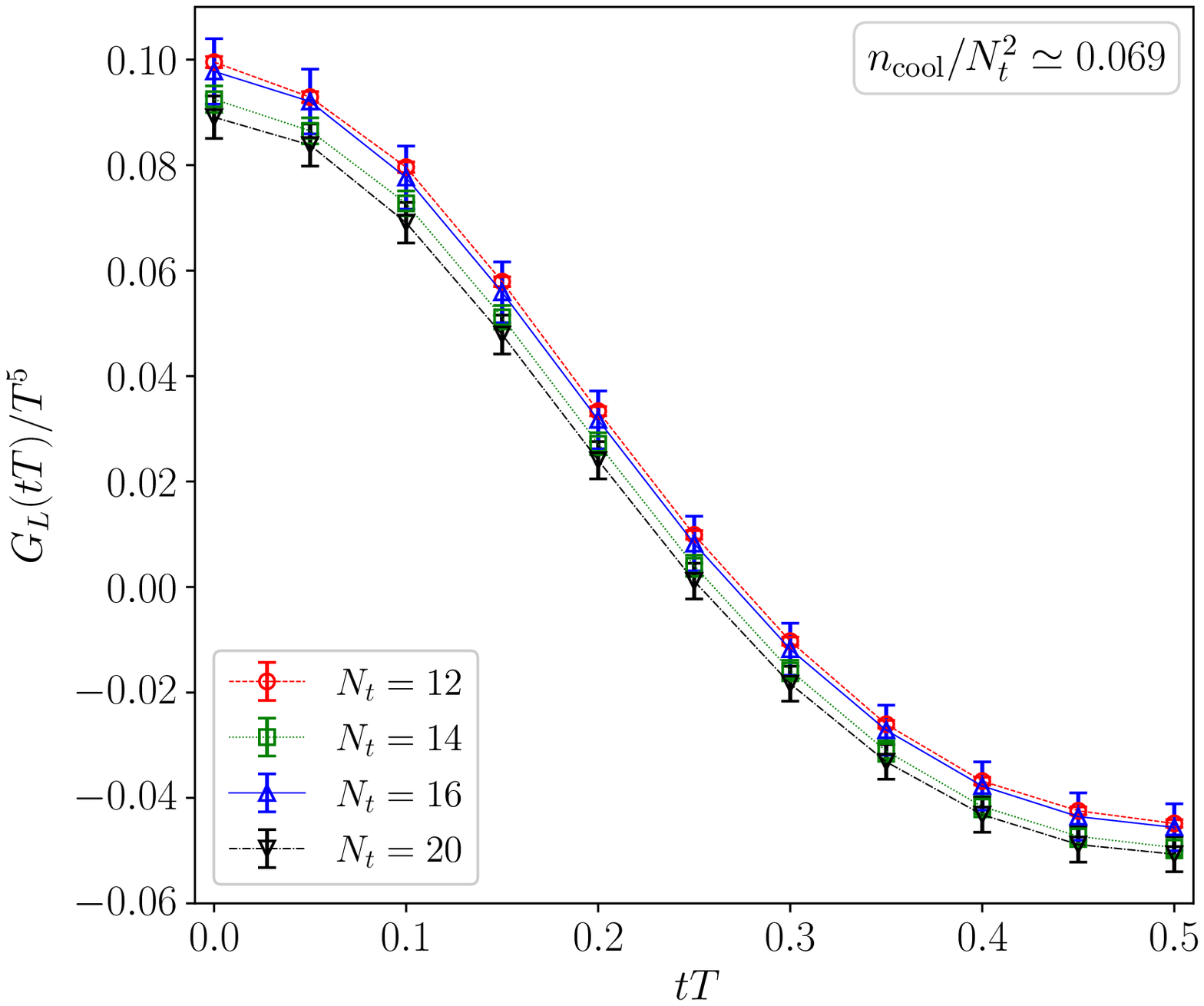}
\includegraphics[scale=0.45]{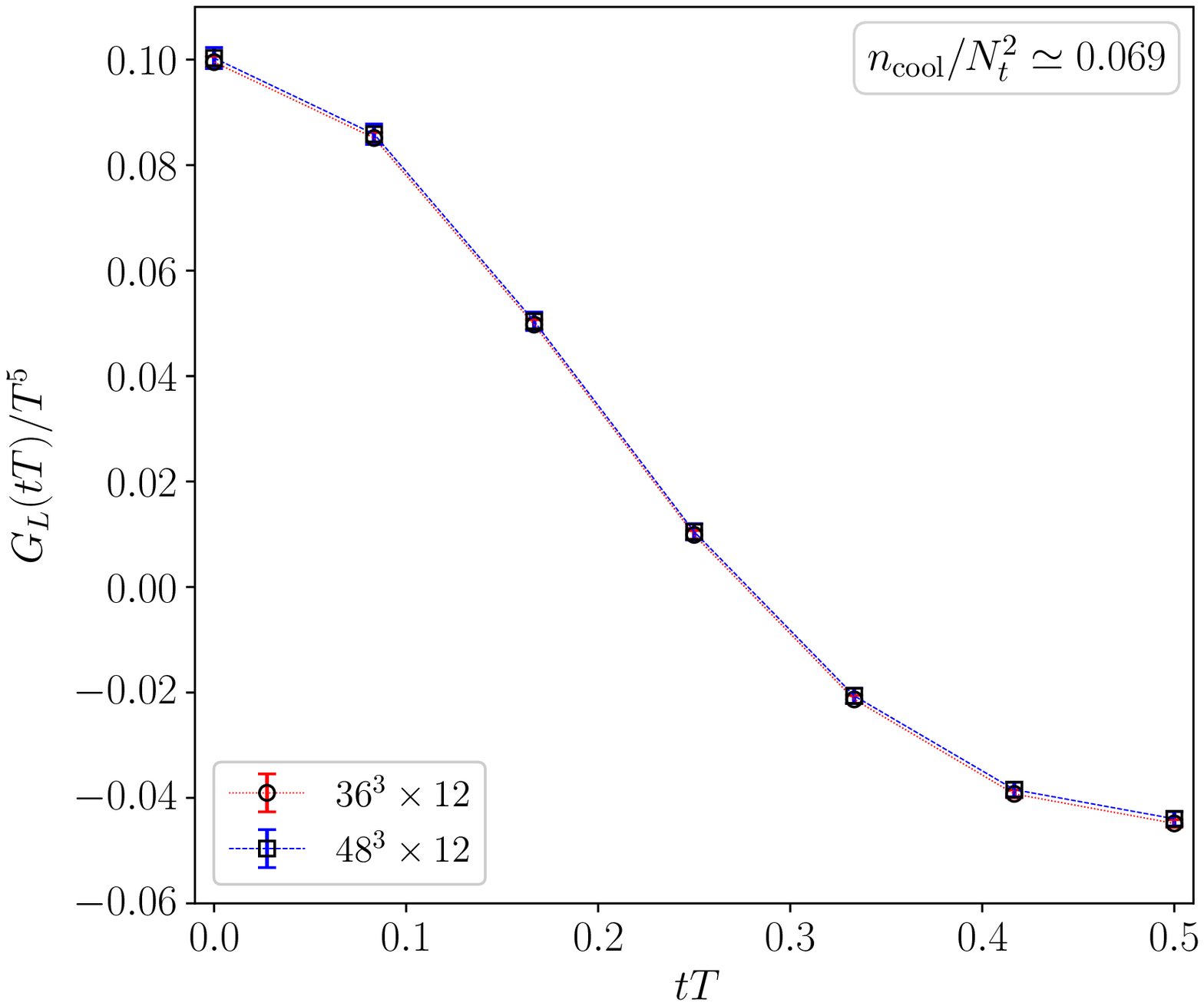}
\caption{Top: determinations of the correlator $G_L(tT)$ for $n_\cool/N_t^2\simeq 0.069$ for all explored values of the lattice spacing. Bottom: comparison of the correlators obtained at $\beta=6.440$ for $n_\cool/N_t^2\simeq 0.069$ on a $36^3 \times 12$ and on a $48^3 \times 12$ lattices. Lines connecting the points have been plotted just to guide the eye.}
\label{fig:tcorr_ex}
\end{figure}

In Fig.~\ref{fig:tcorr_ex}, we show the behavior of the $tT$ and $n_\cool$-interpolated correlators for $n_\cool/N_t^2 \simeq 0.069$ as a function of $tT$ for all explored lattice spacings. Moreover, in Fig.~\ref{fig:tcorr_ex} we also show the comparison between the correlators obtained for $n_\cool/N_t^2 \simeq 0.069 $ for $\beta= 6.440$ on a $36^3 \times 12$ and a $48^3 \times 12$ lattice. Results fall on top of each other, thus we assume that our results obtained on lattices with aspect ratio 3 and spatial extent of $\sim 1.66$~fm do not suffer for significant finite size effects.

To take the continuum limit, we will assume standard $O(a^2) = O(1/N_t^2)$ corrections and we will fit our data for different values of $\beta$ according to the following fit function:
\beq
\begin{aligned}\label{eq:fit_func_cont_lim}
\frac{G_L\left(tT, N_t, \frac{n_\cool}{N_t^2}\right)}{T^5} =&\,\,
\frac{G\left(tT, \frac{n_\cool}{N_t^2}\right)}{T^5} \\&+ c\left(tT, \frac{n_\cool}{N_t^2}\right) \frac{1}{N_t^2} + o\left(\frac{1}{N_t^2}\right),
\end{aligned}
\eeq
where $c$ is a constant factor that, in principle, depends both on the time separation of the sources in the correlator and on the smoothing radius.

\begin{figure}[!htb]
\centering
\includegraphics[scale=0.465]{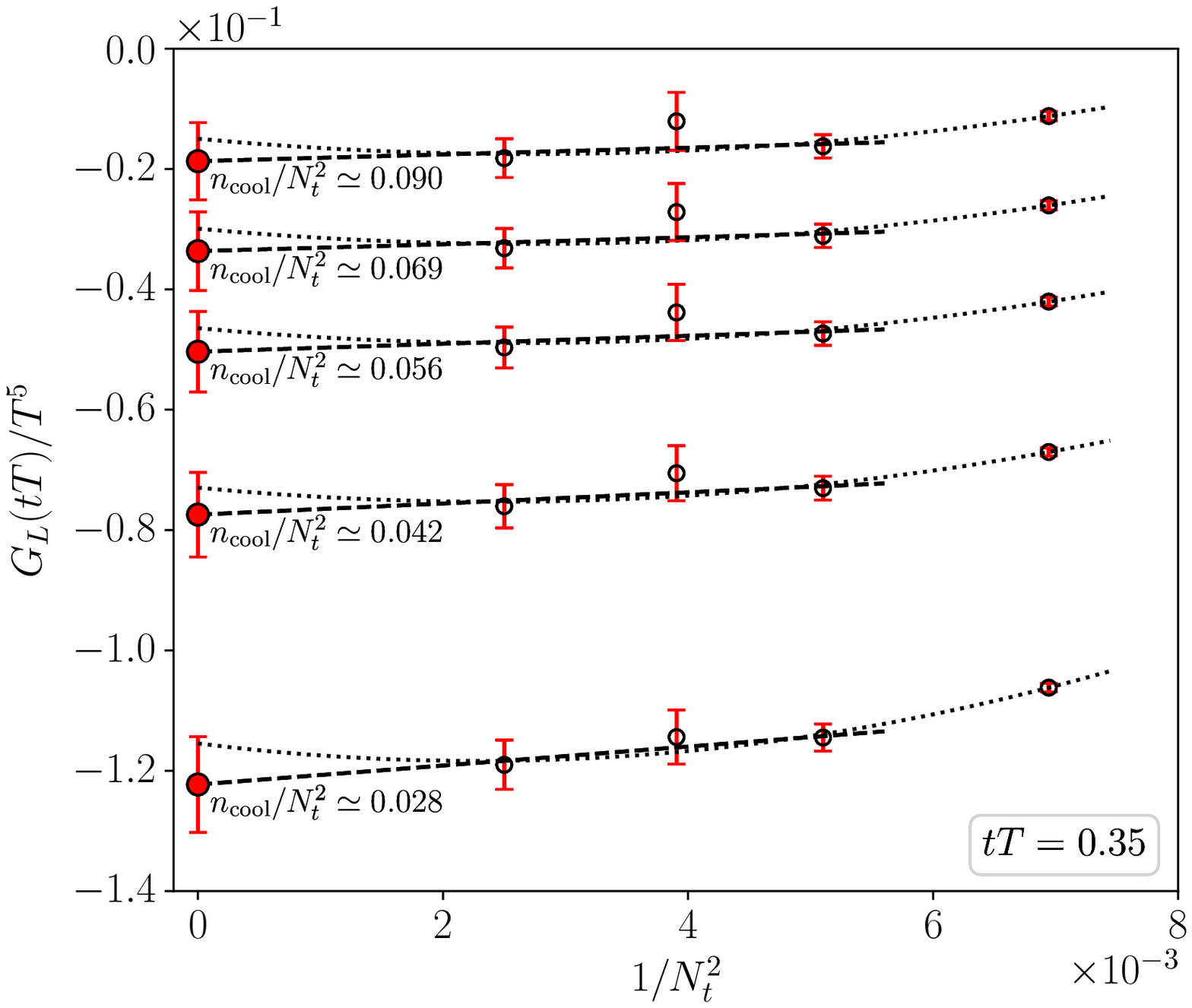}
\includegraphics[scale=0.465]{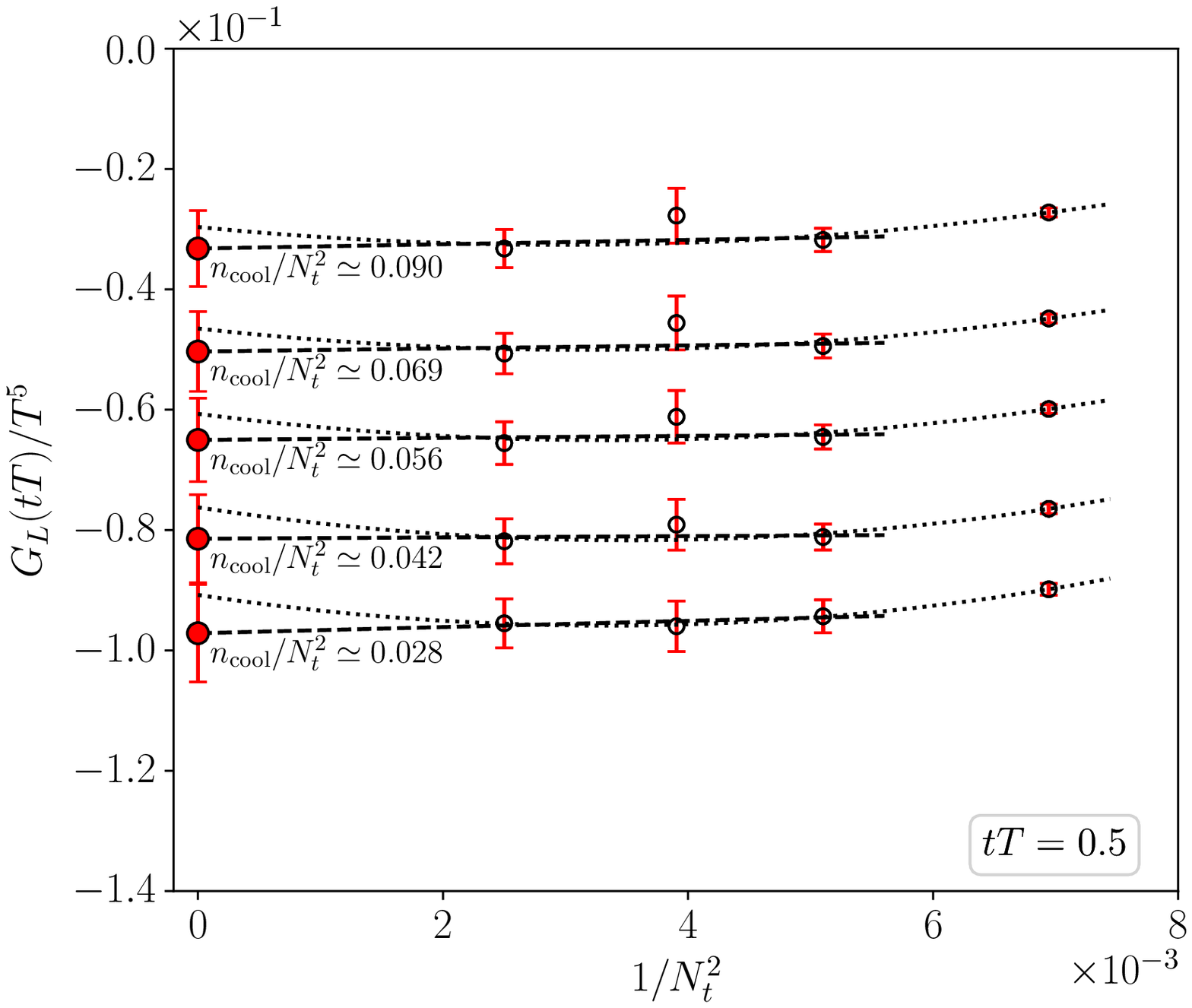}
\caption{Examples of the continuum extrapolation at fixed $n_\cool/N_t^2$ of the correlator for two different values of $tT$.}
\label{fig:tcorr_cont_extr}
\end{figure}

Examples of the continuum limit of $G_L(tT, N_t, n_\cool/N_t^2)$  for two values of $tT$ according to fit function~\eqref{eq:fit_func_cont_lim} are shown in Fig.~\ref{fig:tcorr_cont_extr}. We observe that results at our 3 finest lattice spacings can be reliably fitted with a linear function in $1/N_t^2$. Compatible extrapolations within the errors are obtained fitting all available points and including further $1/N_t^4$ corrections, cf.~Fig.~\ref{fig:tcorr_cont_extr}. Therefore, in what follows we employed the extrapolations obtained with the first fit as our estimates of the continuum limit of the correlator.

Once the correlator is extrapolated towards the continuum limit, there is a residual dependence on the smoothing radius $r_s$. In Ref.~\cite{Altenkort:2020axj} it was shown, using the gradient flow formalism, that the dependence of the continuum-extrapolated correlator is linear in the flow time $\tau_{\mathrm{flow}} \propto r_s^2$. Given that the linear relation $\tau_{\mathrm{flow}}/a^2 = n_{\cool}/3$~\cite{Bonati:2014tqa} holds for the Wilson action in the pure $\SU(3)$ gauge theory, we thus expect to observe a linear dependence on $n_{\cool}/N_t^2$ of our continuum-extrapolated correlator\footnote{See also Refs.~\cite{Bonanno:2022dru,Bonanno:2022hmz}, where a linear behavior on $n_\cool$ is observed in $2d$ $\mathrm{CP}^{N-1}$ models for, respectively, the continuum limit at fixed smoothing radius in physical units of the topological susceptibility $\chi$ and of the topological susceptibility slope $\chi^\prime$.}. Therefore, our final double-extrapolated correlator $G(tT)/T^5$ is obtained from a linear fit in $n_\cool/N_t^2$ according to the fit function:
\beq\label{eq:fit_func_zerocool_lim}
\frac{G\left(tT, \frac{n_\cool}{N_t^2}\right)}{T^5} = \frac{G(tT)}{T^5} + \tilde{c}(tT) \frac{n_\cool}{N_t^2},
\eeq
where $\tilde{c}$ is a constant factor depending on the value of the time separation $tT$.

When performing such zero-cooling extrapolation of $G(tT, n_\cool/N_t^2)$, we fixed the fit range following these prescriptions. For the upper bound, we chose $n_\cool^{(\mathrm{max})}$ in order to ensure that $r_sT < tT$, i.e., cf.~Eq.~\eqref{eq:smooth_rad_cool}:
\beq
\frac{n_\cool^{(\mathrm{max})}}{N_t^2} \lesssim \frac{3}{8} (tT)^2.
\eeq
For our largest time separation $tT=0.5$, we could extend our linear fit region up to $n_\cool/N_t^2 \simeq 0.090$, corresponding, respectively, to $n_\cool \lesssim 13,18,24,37$ for $N_t=12,14,16,20$.

\begin{figure}[!t]
\centering
\includegraphics[scale=0.465]{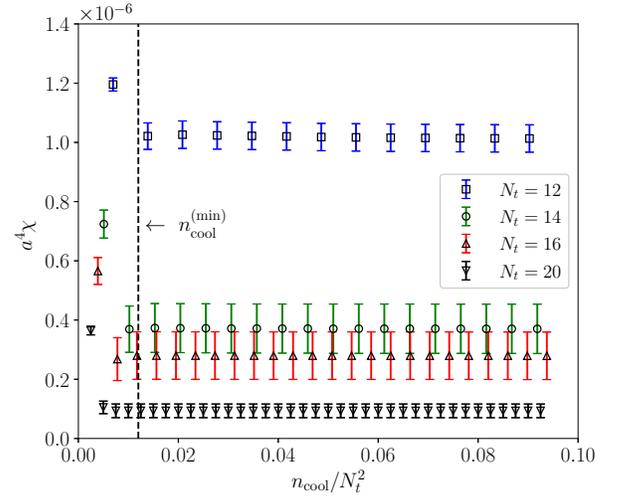}
\caption{Behavior of the topological susceptibility as a function of the number of cooling steps for the four explored lattice spacings. The dashed line denotes the minimum value of $n_\cool/N_t^2$ employed for the zero-cooling extrapolation.}
\label{fig:topsusc_vs_ncool}
\end{figure}

For the lower bound, we choose $n_\cool^{(\mathrm{min})}$ in order to ensure that the topological susceptibility\footnote{The topological susceptibility was computed using the so-called \emph{$\alpha$-rounded} lattice charge, i.e., defining $Q=\mathrm{round}[\alpha Q_L(n_\cool)]$, where $Q_L(n_\cool)$ is the definition in Eq.~\eqref{eq:topcharge_profile} computed after $n_\cool$ cooling steps and $\alpha$ is found by minimizing the mean squared difference between $\alpha Q_L(n_\cool)$ and $\mathrm{round}[\alpha Q_L(n_\cool)]$~\cite{DelDebbio:2002xa,Bonati:2015sqt}.} $a^4 \chi = \braket{Q^2}/(N_s^3 N_t)$ has reached a plateau (as a function of $n_\cool/N_t^2$) for all the explored values of $\beta$, cf.~Fig.~\ref{fig:topsusc_vs_ncool}. In our case, it turns out that $n_\cool/N_t^2 = 0.012$ is a reasonable lower bound, corresponding, respectively, to $n_\cool \gtrsim 1,2,3,4$ for $N_t=12,14,16,20$.

These prescriptions were chosen to ensure that we did enough cooling so as to correctly identify the correct topological charge for all the lattice configurations, but at the same time that we did not do too much cooling so as to make the sources in the correlator overlap onto each other.

However, a drawback of this procedure is that, when $tT$ approaches 0, the fit range becomes narrower and narrower, eventually closing. As a matter of fact, for time separations $tT \le 0.2$ we could not perform a reliable zero-cooling extrapolation. Therefore, we could only compute the double-extrapolated correlator for $tT > 0.2$.

\begin{figure}[!t]
\centering
\includegraphics[scale=0.5]{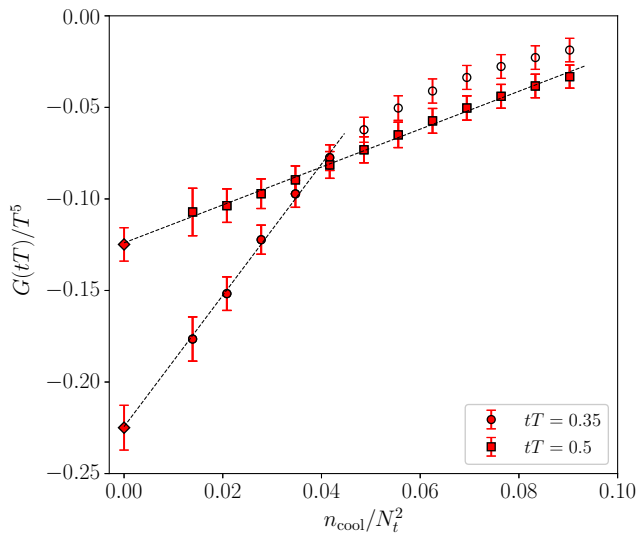}
\caption{Examples of the zero-cooling extrapolation of the correlator $G(tT,n_\cool/N_t^2)$ for two different values of $tT$.}
\label{fig:tcorr_zerocool_extr}
\end{figure}

\begin{figure}[!t]
\centering
\includegraphics[scale=0.5]{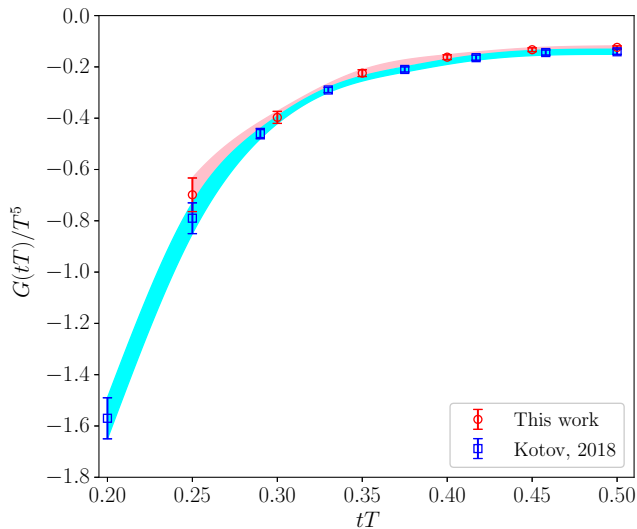}
\caption{Comparison of the results for the double extrapolated correlator $G(tT)/T^5$ obtained in this work with those reported in Ref.~\cite{Kotov:2018aaa} at the same temperature and using the gradient flow as smoothing method.}
\label{fig:comp_Kotov}
\end{figure}

In Fig.~\ref{fig:tcorr_zerocool_extr} we show examples of the zero-cooling extrapolation for two values of $tT$, while in Fig.~\ref{fig:comp_Kotov} we show our complete double-extrapolated correlator $G(tT)/T^5$. Our final correlator turns out to be negative in all cases as expected, and in overall good agreement with the double-extrapolated correlator obtained for the same temperature in Ref.~\cite{Kotov:2018aaa}, where the gradient flow was used as smoothing method to define the lattice topological charge density.\\

\begin{figure}[!t]
\centering
\includegraphics[scale=0.5]{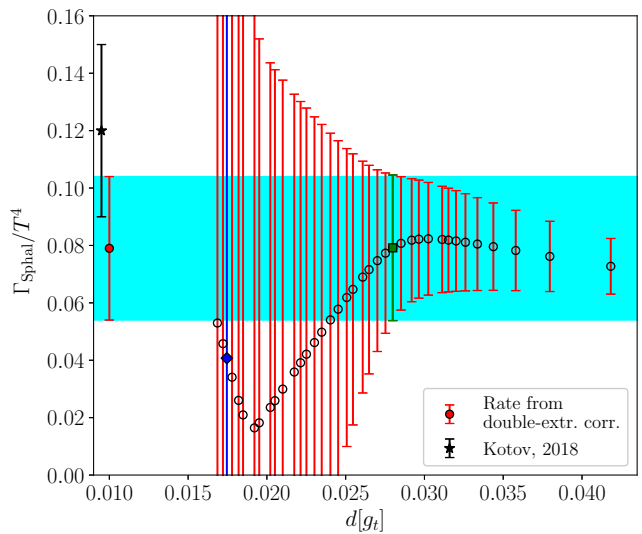}
\caption{Results for the rate $\Gamma_\sphal$ as a function of $d[g_t]$, defined in Eq.~\eqref{eq:def_d}, extracted from the double extrapolated correlator. The square and diamond points represent, respectively, our choices for $\lambda_1$ and $\lambda_2$, see discussion below Eq.~\eqref{eq:def_d} for more details. The full point and the shaded area represent our final result for $\Gamma_\sphal$.}
\label{fig:rate_vs_lambda_double_extr}
\end{figure}

We can now invert our double-extrapolated correlator to obtain the sphaleron rate using the inversion method outlined in Sec.~\ref{sec:backus-gilbert}. Obtained results are reported in Fig.~\ref{fig:rate_vs_lambda_double_extr} as a function of the parameter $d[g_t](\lambda)$.

As it can be appreciated, the fact that it is possible to reconstruct the double-extrapolated correlator only for $tT>0.2$ leads to a noisy reconstruction of the sphaleron rate, which suffers from quite large errors, especially for small values of $d[g_t]$. We quote as our final result:
\beq\label{eq:final_res_rate_form_double_extr_corr}
\frac{\Gamma_\sphal}{T^4} = 0.079(25), \qquad T \simeq 1.24~T_c.
\eeq

This result, depicted as a round point and as a uniform shaded area in Fig.~\ref{fig:rate_vs_lambda_double_extr}, is indeed compatible with all the other results for the rate at smaller/larger values of $d[g_t](\lambda)$, and any variation observed in the central value of the rate as a function of $d$ is much smaller than the errors on the points, signalling that our reconstruction is stable as a function of the regulator $\lambda$ defining our minimized functional in Eq.~\eqref{eq:backus_gilbert_functional_tot}.

Let us now comment how our result in Eq.~\eqref{eq:final_res_rate_form_double_extr_corr} compares with the result of Ref.~\cite{Kotov:2018aaa}. Although the central value is $\sim 33\%$ smaller, our number is compatible within errors with the one reported in Ref.~\cite{Kotov:2018aaa} for the rate at the same $T$, $\Gamma_\sphal/T^4 = 0.12(3)$, obtained applying the standard Backus--Gilbert method~\cite{BackusGilbert1968:aaa}.

\FloatBarrier

\subsection{Double extrapolation of the sphaleron rate}\label{sec:double_extr_rate}

In this section, we follow a different strategy to compute the sphaleron rate; namely, we extract $\Gamma_{\sphal,L}$ from the correlators $G_L(tT)$ obtained at finite lattice spacing as a function of $n_\cool$, using the same inversion method of Sec.~\ref{sec:backus-gilbert}, with the aim of postponing the double-extrapolation of the correlator directly onto the rate itself. A first bonus feature of this approach is that no time interpolation of the correlators is now needed in the double-extrapolation procedure.

\begin{figure}[!t]
\centering
\includegraphics[scale=0.48]{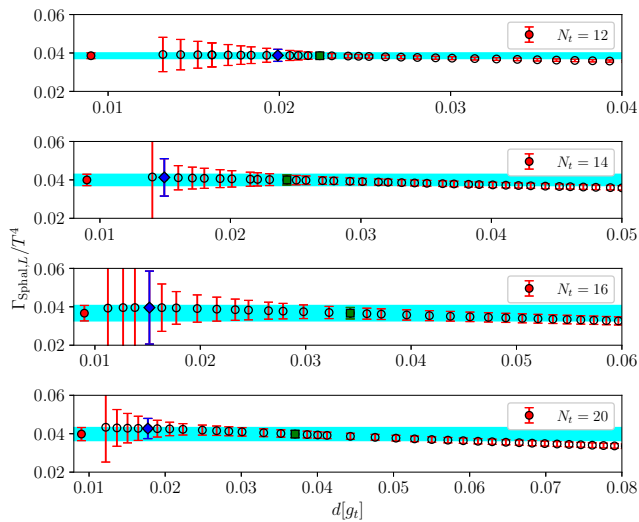}
\caption{Results for the rate $\Gamma_{\sphal,L}$ as a function of $d[g_t]$, defined in Eq.~\eqref{eq:def_d}, extracted from the finite lattice spacing correlators and for, respectively, $n_\cool = 10, 14, 18, 28$ for $N_t=12,14,16,20$. Square and diamond points represent, respectively, our choices for $\lambda_1$ and $\lambda_2$, see discussion below Eq.~\eqref{eq:def_d} for more details. The full points and the shaded areas represent our final results for $\Gamma_{\sphal,L}$.}
\label{fig:ex_rates}
\end{figure}

\begin{figure}[!t]
\centering
\includegraphics[scale=0.47]{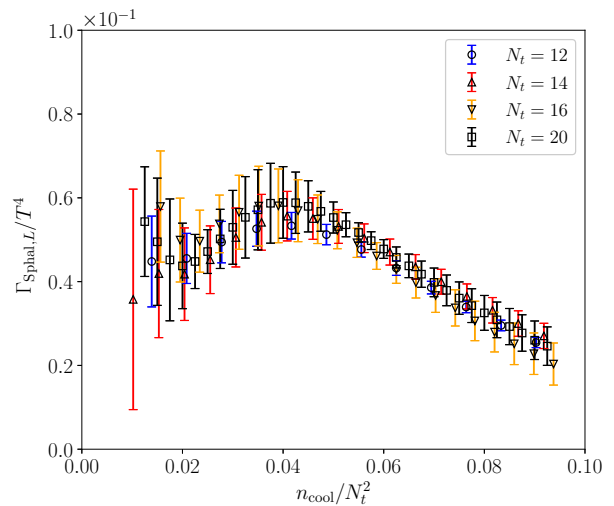}
\caption{Results for $\Gamma_{\sphal,L}/T^4$ as a function of $n_\cool$ for all the explored lattice spacings.}
\label{fig:comp_rates}
\end{figure}

In Fig.~\ref{fig:ex_rates} we show examples of the results obtained from the modified Backus--Gilbert for all available values of $N_t$ and for approximately the same value of $n_\cool/N_t^2$. As it can be seen, the reconstruction of the sphaleron rate from lattice correlators is more stable compared to the one obtained from the double extrapolated one, cf.~Fig.~\ref{fig:rate_vs_lambda_double_extr}. In Fig.~\ref{fig:comp_rates} we collect our results for the rate at finite lattice spacing as a function of $n_\cool$ for every $N_t$ explored.

Once $\Gamma_{\sphal,L}(a,n_\cool)$ is determined, we can perform the continuum limit at fixed smoothing radius $(r_sT)^2\propto n_\cool/N_t^2$ according to the fit function:
\beq
\begin{aligned}
\frac{\Gamma_{\sphal,L}}{T^4}\left(N_t, \frac{n_\cool}{N_t^2}\right) =&\,\, \frac{\Gamma_\sphal}{T^4}\left(\frac{n_\cool}{N_t^2}\right) \\ &+ k\left(\frac{n_\cool}{N_t^2}\right) \frac{1}{N_t^2},
\end{aligned}
\eeq
where $k$ is a constant factor depending on the value of $n_\cool/N_t^2$.

\begin{figure}[!t]
\centering
\includegraphics[scale=0.47]{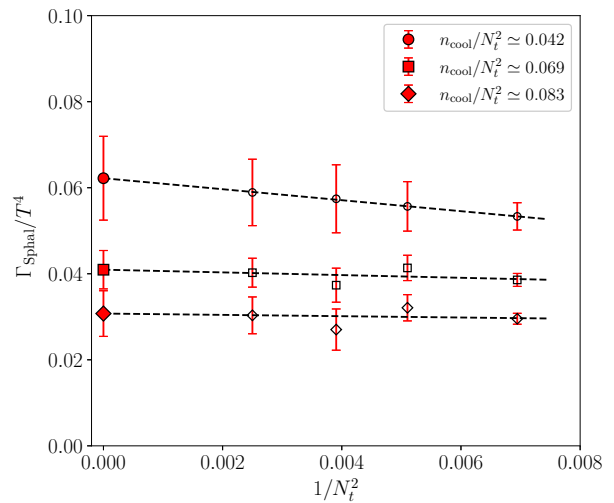}
\caption{Continuum extrapolation of the sphaleron rate at fixed smoothing radius $(r_sT)^2 \propto n_\cool/N_t^2$ for a few values of $n_\cool/N_t^2$.}
\label{fig:rate_cont_limit}
\end{figure}

\begin{figure}[!t]
\centering
\includegraphics[scale=0.48]{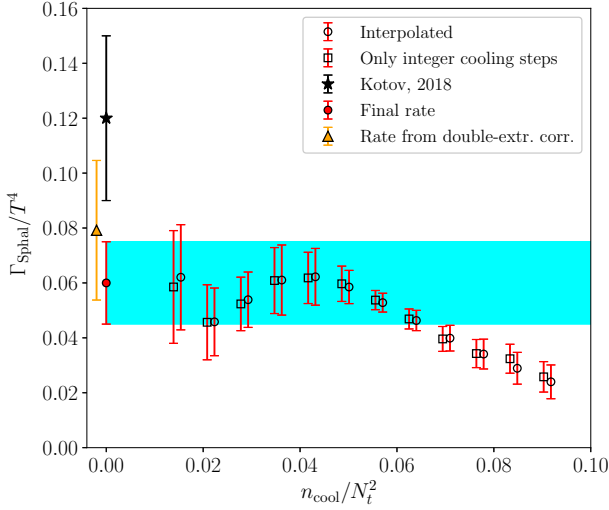}
\caption{Dependence of the continuum-extrapolated sphaleron rate on the smoothing radius $(r_s T)^2 \propto n_\cool/N_t^2$. The full round point and  the shaded area represent our final result for $\Gamma_\sphal/T^4$. The full triangle and starred points represent, respectively, the rate obtained from the inversion of the double-extrapolated correlator, and the one computed in Ref.~\cite{Kotov:2018aaa} at the same temperature, but adopting the standard Backus--Gilbert method and using the gradient flow as smoothing method.}
\label{fig:rate_zerocool}
\end{figure}

Also in this case, in order to keep $n_\cool/N_t^2$ fixed, we have performed a spline cubic interpolation of our results of $\Gamma_{\sphal,L}/T^4$ as a function of $n_\cool$. Examples of continuum extrapolations of $\Gamma_{\sphal,L}$ for a few values of $n_\cool/N_t^2$ are shown in Fig.~\ref{fig:rate_cont_limit}. Interestingly enough, unlike what has been observed for the topological charge density correlator, we observe a very mild dependence of the sphaleron rate on the lattice spacing. As a matter of fact, it is possible to obtain an excellent best fit of our data with a linear function in $1/N_t^2$ using all available values of $N_t$, and results obtained restricting such fit to our three finest lattice spacings turn out in excellent agreement within the errors. Our continuum extrapolations of $\Gamma_\sphal$ as a function of $n_\cool/N_t^2$ are shown in Fig.~\ref{fig:rate_zerocool}.

Before discussing further our results for the sphaleron rate, let us first make a comment about the $n_\cool$ interpolation. From Fig.~\ref{fig:comp_rates}, we observe that the $n_\cool$ dependence of $\Gamma_{\sphal,L}$ is pretty mild, in particular for small values of $n_\cool$, thus, it is reasonable to believe that the rate will vary only little upon the $n_\cool$ interpolation. To check this assumption, we have also performed our continuum extrapolation at fixed smoothing radius in the following way: given a value of $n_\cool$ for the lattice with the smallest temporal extent $N_t=12$, the corresponding (integer) value $n_\cool'$ for another temporal extent $N_t$ is given by $n_\cool' = \mathrm{round} \left[ n_\cool (N_t/12)^2 \right]$ (cf.~Eq.~\eqref{eq:smooth_rad_cool}) where $\mathrm{round}[x]$ denotes the closest integer to $x$. Results obtained with this approximation are shown in Fig.~\ref{fig:rate_zerocool} as square points. As it can be appreciated, no difference is observed in the final continuum extrapolation for the sphaleron rate compared to the ones obtained interpolating in $n_\cool$ (round points). Therefore, we can conclude that, although in principle being a better approximation to keep $n_\cool /N_t^2$ fixed among different lattices with different temporal extents $N_t$, in the end not even the $n_\cool$ interpolation is needed with this approach.

Let us now discuss the dependence of our results in Fig.~\ref{fig:rate_zerocool} on the cooling radius. We observe that $\Gamma_\sphal$ does not show a sizeable dependence on the smoothing radius for small enough values of $n_\cool/N_t^2$, and in particular it approaches a plateau for $n_\cool/N_t^2 \lesssim 0.045$, see Fig.~\ref{fig:rate_zerocool}. As already discussed in the Introduction, this behavior is perfectly reasonable, since smoothing is expected to only modify the high-frequency components of the spectral density. Thus, being $\Gamma_\sphal$ related to the zero-frequency limit of $\rho(\omega)/\omega$, one can expect this quantity to become insensitive to the value of the smoothing radius, as long as the UV cut-off introduced by the smoothing radius is sufficiently separated from the typical IR scale of the relevant topological fluctuations contributing to the sphaleron rate.

\begin{figure}[!t]
\centering
\includegraphics[scale=0.435]{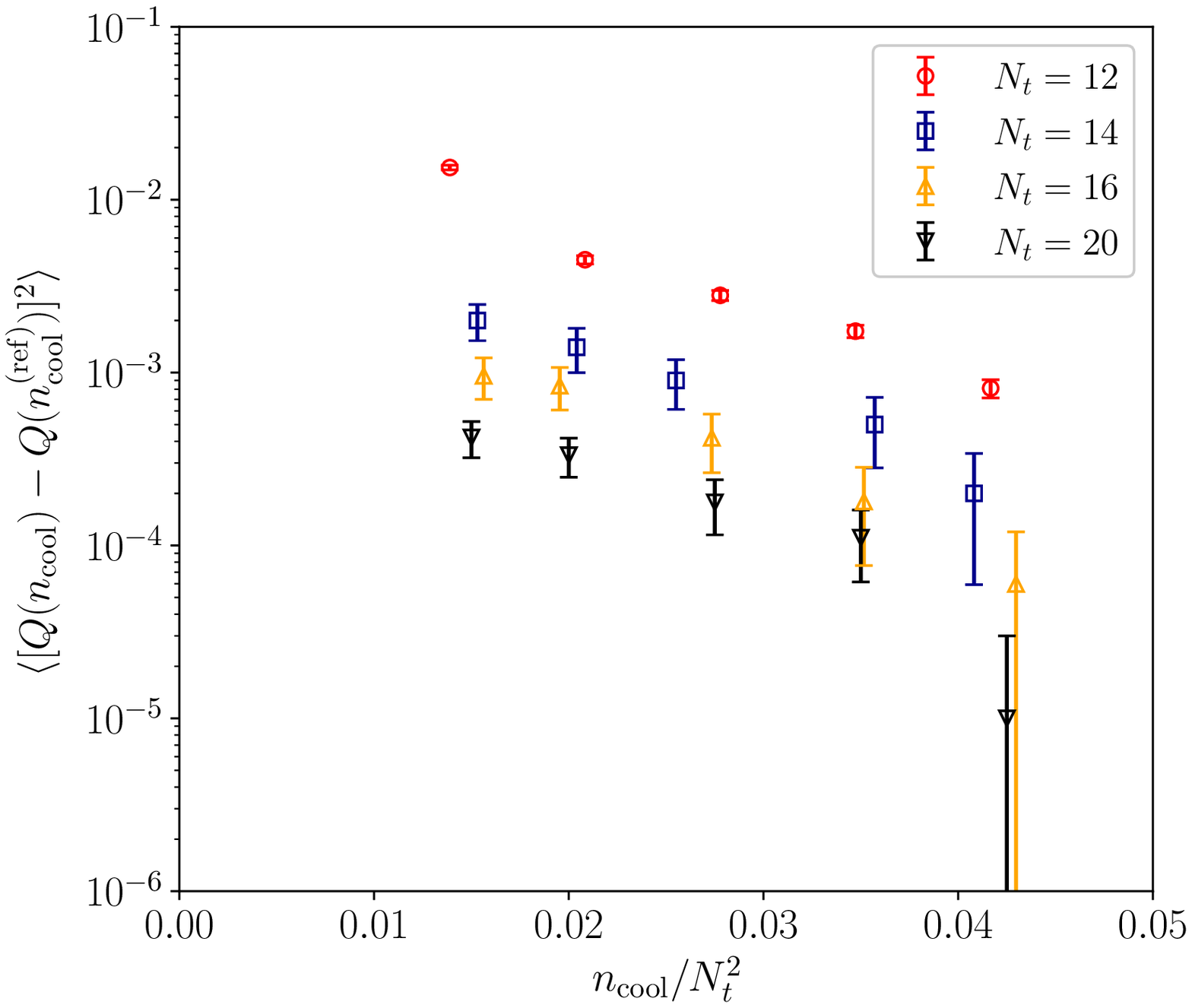}
\includegraphics[scale=0.435]{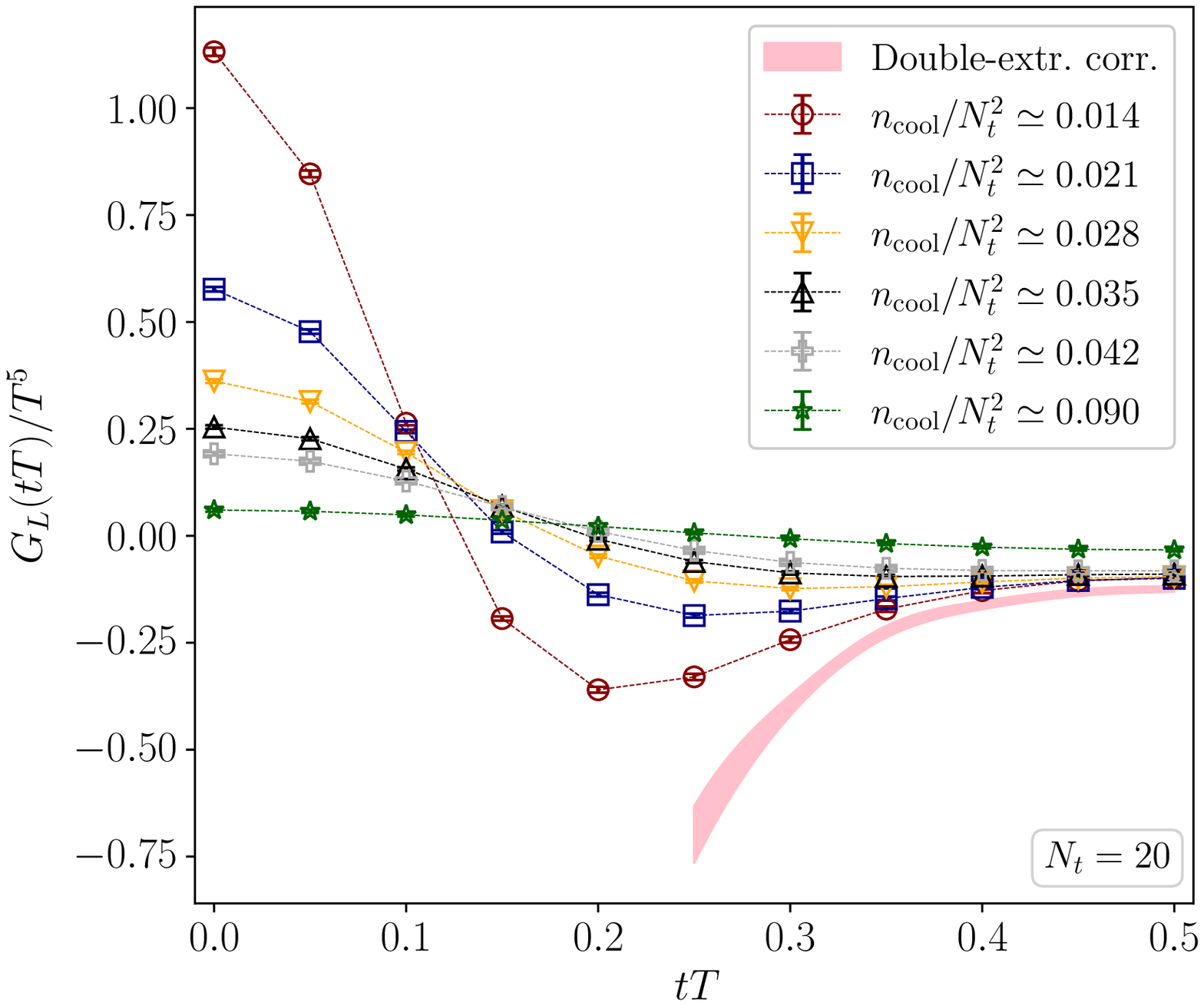}
\caption{Top panel: study of the stability of the topological charge, assessed from the mean-square-difference between the determinations of $Q$ after $n_\cool$ cooling steps with respect to $Q$ determined for a reference number of cooling steps $n_\cool^{(\mathrm{ref})}$. In all cases $n_\cool$ was varied within $0.015 \lesssim \Gamma_\sphal \lesssim 0.045$, and $n_\cool^{(\mathrm{ref})} /N_t^2 \sim 0.045$, corresponding to: $[2,6]$ for $N_t=12$ ($n_\cool^{(\mathrm{ref})} = 7$), $[3,8]$ for $N_t=14$ ($n_\cool^{(\mathrm{ref})} = 10$), $[4,11]$ for $N_t=16$ ($n_\cool^{(\mathrm{ref})} = 12$) and $[6,17]$ for $N_t=20$ ($n_\cool^{(\mathrm{ref})} = 19$). Bottom panel: Comparison of the lattice correlators obtained for $N_t=20$ and for smoothing radii chosen in the range $0.015\lesssim n_\cool/N_t^2 \lesssim 0.04$, corresponding to the plateau observed for our continuum results for $\Gamma_\sphal$ as a function of $n_\cool/N_t^2$, cf.~Fig.~\ref{fig:rate_zerocool}. For the sake of comparison, we also show the correlator obtained for $n_\cool/N_t^2 \simeq 0.09$, and the double-extrapolated correlator computed in Sec.~\ref{sec:double_extr_corr}, depicted as a uniform shaded area.}
\label{fig:comp_tcorrs_ncool_Nt_20}
\end{figure}

In order to check that this is what is actually happening, it is interesting to take a look at the behavior of the correlator of the topological charge density as a function of the number of cooling steps. In Fig.~\ref{fig:comp_tcorrs_ncool_Nt_20} (bottom panel) we compare correlators obtained for $N_t=20$ and for smoothing radii chosen within the range $0.015 \lesssim n_\cool/N_t^2 \lesssim 0.045$, corresponding to the plateau observed in Fig.~\ref{fig:rate_zerocool}. As it can be appreciated, varying the smoothing radius changes sizeably the short-distance behavior of the correlator, as expected, while it has a much smaller impact on the long-distance tail of the correlator, leading to very small variations for $tT\ge0.4$. On the other hand, the correlator obtained for the largest smoothing radius considered in our study, corresponding to $n_\cool/N_t^2 \simeq 0.09$, sensibly deviates from those obtained at smaller smoothing radii even up to $tT=0.5$, cf.~Fig.~\ref{fig:comp_tcorrs_ncool_Nt_20}. In this case, the smoothing radius is so large that it has a visible impact on the long-distance behavior of $G(t)$, and this reflects in a smaller value of the sphaleron rate, cf.~Fig.~\ref{fig:rate_zerocool}.

With the purpose of verifying the reliability of our determinations of the correlators in the range of smoothing radii where $\Gamma_\sphal$ exhibits a plateau, we also checked that, in the same range, the determination of the topological background is already well defined and stable.
The result of this study is shown in the top panel of Fig.~\ref{fig:comp_tcorrs_ncool_Nt_20}, where we show the quantity:
\beq\label{eq:diff_Q_vs_ncool}
\left\langle\left[ Q(n_\cool) - Q(n_\cool^{(\mathrm{ref})}) \right]^2\right\rangle
\eeq
as a function of $n_\cool$. Here, $n_\cool^{(\mathrm{ref})}$ is defined as the number of cooling steps corresponding to approximately $n_\cool/N_t^2 \simeq 0.045$, i.e., the upper bound of the range we are interested in, while $n_\cool$ varies down to values corresponding approximately to $n_\cool/N_t^2 \simeq 0.015$, i.e., the lower bound of the range we are interested in. As it can be observed from Fig.~\ref{fig:comp_tcorrs_ncool_Nt_20}, the quantity in Eq.~\eqref{eq:diff_Q_vs_ncool} is in the worst case $\sim 10^{-2}$, meaning that, upon varying the number of cooling steps within the range $0.015 \lesssim n_\cool/N_t^2 \lesssim 0.045 $, the fraction of configurations whose assigned topological charge varies is at most the $\sim 1\%$ of the whole ensemble (as in all the explored cases we observed $\vert \Delta Q \vert = 0,1$). Therefore, our determinations of the correlator in the 
range of smoothing radii where $\Gamma_\sphal$ is flat, are taken in a regime where IR topological fluctuations are well defined.

In the light of this discussion, in this case we do not perform any zero-cooling extrapolation, and simply take the value of the plateau exhibited by the sphaleron rate for small cooling radii (corresponding to the range $0.015 \lesssim n_\cool/N_t^2 \lesssim 0.045$) as our final result for $\Gamma_\sphal$. Such result is depicted in Fig.~\ref{fig:rate_zerocool} as a shaded uniform area and as a full round point in $n_\cool/N_t^2=0$, and corresponds to:
\beq\label{eq:final_result_rate}
\frac{\Gamma_\sphal}{T^4} = 0.060(15), \qquad T \simeq 1.24~T_c,
\eeq
where the central value and the uncertainty are chosen taking into account the central values of the points on the plateau,
their error bars and the residual observed variability.

We would like to stress that considering instead a zero-smoothing extrapolation, which also involves larger values of 
$n_\cool/N_t^2$, would not be well justified within our approach, in view of the sizable distortions of the 
correlator affecting such values and of the absence of a sound theoretical framework~\cite{Altenkort:2020fgs}
to perform such extrapolation. Looking for a plateau as a function of $n_\cool/N_t^2$ is instead,
as long as such plateau is actually observed and well defined, more solid and sound.

The result in Eq.~\eqref{eq:final_result_rate} turns out to be compatible with the one found from the inversion of the double extrapolated correlator illustrated in Sec.~\ref{sec:double_extr_corr}, $\Gamma_\sphal/T^4=0.079(25)$, but has a smaller relative uncertainty. Moreover, also this result points towards a smaller central value for the sphaleron rate compared to the one reported in Ref.~\cite{Kotov:2018aaa} at the same temperature, $\Gamma_\sphal/T^4=0.12(3)$, even if it is still compatible with it within less than two standard deviations.

We can also compare our results with the recent determination of Ref.~\cite{BarrosoMancha:2022mbj}, where a completely different strategy to compute $\Gamma_\sphal$ from quenched lattice simulations was pursued. The smallest temperature explored in that work is $T \simeq 1.3~T_c$, which is very close but not exactly equal to the one studied here, $T \simeq 1.24~T_c$. However, our result turns out to be in perfect agreement with the one reported in that paper at that temperature: $\Gamma_\sphal/T^4 = 0.061(2)$.

\section{Conclusions}\label{sec:conclu}

In this work we have computed the sphaleron rate $\Gamma_\sphal$ in quenched QCD for a temperature $T\simeq 1.24~T_c \simeq 357$~MeV from lattice numerical Monte Carlo simulations using the modified Backus--Gilbert method proposed by the Rome group to invert the integral relation between the Euclidean topological charge density time correlator and the spectral density, whose zero-frequency limit is directly related to $\Gamma_\sphal$.

We have followed two strategies. The first one is similar to what has been already done in the past, namely, we have performed a double extrapolation of the topological charge density correlator (continuum limit at fixed smoothing radius in physical units followed by zero-smoothing limit) and then extracted the rate from the inversion of such double-extrapolated correlator. The second method, instead, consists in extracting the rate directly from the inversion of finite-lattice-spacing correlators, in order to postpone the double extrapolation directly on the rate itself.

The two methods give consistent results, but we find that the second is preferable for various reasons. First, it eliminates both the need of interpolating in $tT$ (as the rate is extracted from finite lattice spacing correlators) and in $n_\cool$ (as the rate depends very mildly on the smoothing radius, so that no difference is observed upon interpolating our results for the rate in $n_\cool$, rather than just taking the result for the integer $n_\cool$ closest to the reference smoothing radius). Second, the inversion to reconstruct $\rho(\omega)$ is found to be less noisy compared to the one performed on the double-extrapolated correlator. Finally, we find that the rate is affected by smaller lattice artifacts, and that it is practically insensitive to the value of the smoothing radius for small enough values of $n_\cool$. In the end, thus, the second strategy turns out to be simpler and computationally cheaper, and finally yields a smaller error compared to the first one.

As we have already discussed in the Introduction, one should be careful in applying the second method, since according to 
perturbative arguments~\cite{Altenkort:2020fgs} the integral relation in Eq.~\eqref{eq:rho_def} could be distorted when 
considering the correlator at finite smoothing radius. However, as we have argued above, this problem is expected 
to be less relevant to the $\omega \to 0$ regime involved in the sphaleron rate. As an effective way to check that this 
is indeed the case, we have verified the existence of an extended range of values of the smoothing radius
in which the continuum extrapolated sphaleron rate is practically constant within errors, then taking our 
final determination for the rate from this plateau. We consider the existence of this plateau, over which 
also the topological background is stable, as a solid, even if heuristic, evidence for
a well defined separation between the ultraviolet (UV) cut-off scale and the physical scale of fluctuations relevant 
to the sphaleron rate, which makes our approach justified; the same conclusion is reached also looking at the long-distance tails of the correlators, that appear to be practically unaffected by cooling in the same range where we observe a plateau for the sphaleron rate.

Therefore, in conclusion, while the first standard method, based on the double-extrapolated correlator, is surely better founded 
from a theoretical point of view, it is affected by uncertainties which could make it of difficult application in
contexts, like full QCD with physical quark masses, where the increased computational demand makes statistics 
significantly poorer compared to the quenched case. The second method that we have proposed, instead, even if justified 
a posteriori based on the observation of a well defined plateau for the sphaleron rate as a function of the smoothing radius, 
provides a more precise probe, which could reveal useful in applications to the case of full QCD.

We find our final result for the rate, quoted in Eq.~\eqref{eq:final_result_rate}, to be smaller but compatible within the errors with the one reported in Ref.~\cite{Kotov:2018aaa} for the same temperature, which was obtained inverting the double-extrapolated correlator, but using the gradient flow as smoothing method and using the standard Backus--Gilbert inversion technique to compute $\Gamma_\sphal$. We stress however that the possible (mild) tension is likely not related to the different smoothing procedure, since we also find that our double-extrapolated correlator is in perfect agreement with the one computed in Ref.~\cite{Kotov:2018aaa} at the same $T$.
Finally, perfect compatibility is found with the result obtained for the sphaleron rate at $T\simeq 1.3~T_c$ in Ref.~\cite{BarrosoMancha:2022mbj}, where a completely different method to extract the rate was pursued (based on the computation of the susceptibility of the so-called ``sphaleron topological charge'').

Our present results can be considered as a basis for a future application of the new strategy proposed in this paper 
to the computation of the sphaleron rate in full QCD at finite temperature, being this quantity of great interest both for studying the properties of the quark-gluon plasma and for obtaining intriguing predictions about axion phenomenology.

\acknowledgements
It is a pleasure to thank Giuseppe Gagliardi, Vittorio Lubicz, Francesco Sanfilippo and Giovanni Villadoro for useful discussions. The work of Claudio Bonanno is supported by the Spanish Research Agency (Agencia Estatal de Investigación) through the grant IFT Centro de Excelencia Severo Ochoa CEX2020-001007-S and, partially, by grant PID2021-127526NB-I00, both funded by MCIN/AEI/10.13039/501100011033. Claudio Bonanno also acknowledges support from the project H2020-MSCAITN-2018-813942 (EuroPLEx) and the EU Horizon 2020 research and innovation programme, STRONG-2020 project, under grant agreement No 824093. Numerical simulations have been performed on the \texttt{MARCONI} and \texttt{Marconi100} machines at CINECA, based on the agreement between INFN and CINECA, under projects INF22\_npqcd and INF23\_npqcd.

\appendix
\section{Scale setting}\label{app:scale_setting}

In this work the lattice spacing was determined in units of the Sommer scale $r_0$ using the $\sim 1\%$ determinations of $a(\beta)/r_0$ of Ref.~\cite{Necco:2001xg}, which allowed us to define a LCP where the volume and the temperature are kept fixed at the percent level.

In order to check for possible systematics in such procedure, we also performed a different scale setting according to the results of Ref.~\cite{Francis:2015lha}, which employs a different parameterization to interpolate lattice determinations of $a(\beta)/r_0$. Although these two scale settings are not completely independent, as~\cite{Francis:2015lha} uses lattice spacing results of~\cite{Necco:2001xg} for $\beta < 6.2$ and, most importantly, for $\beta > 6.5$ (the range where the three finest lattice spacings employed here fall), it is still worth checking that setting the scale in a different way gives agreeing results within the typical error on $a/r_0$.

\begin{figure}
\centering
\includegraphics[scale=0.5]{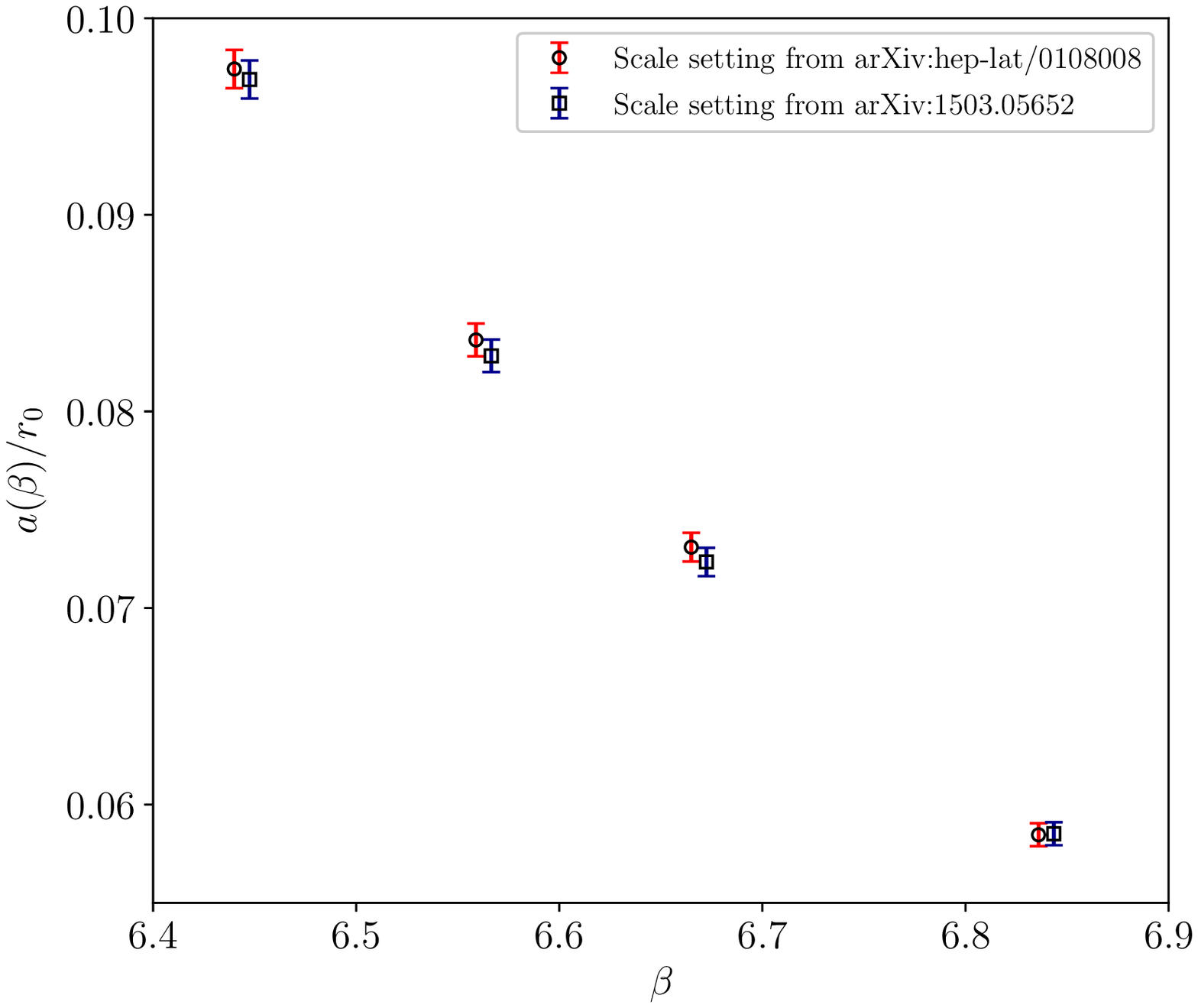}
\includegraphics[scale=0.5]{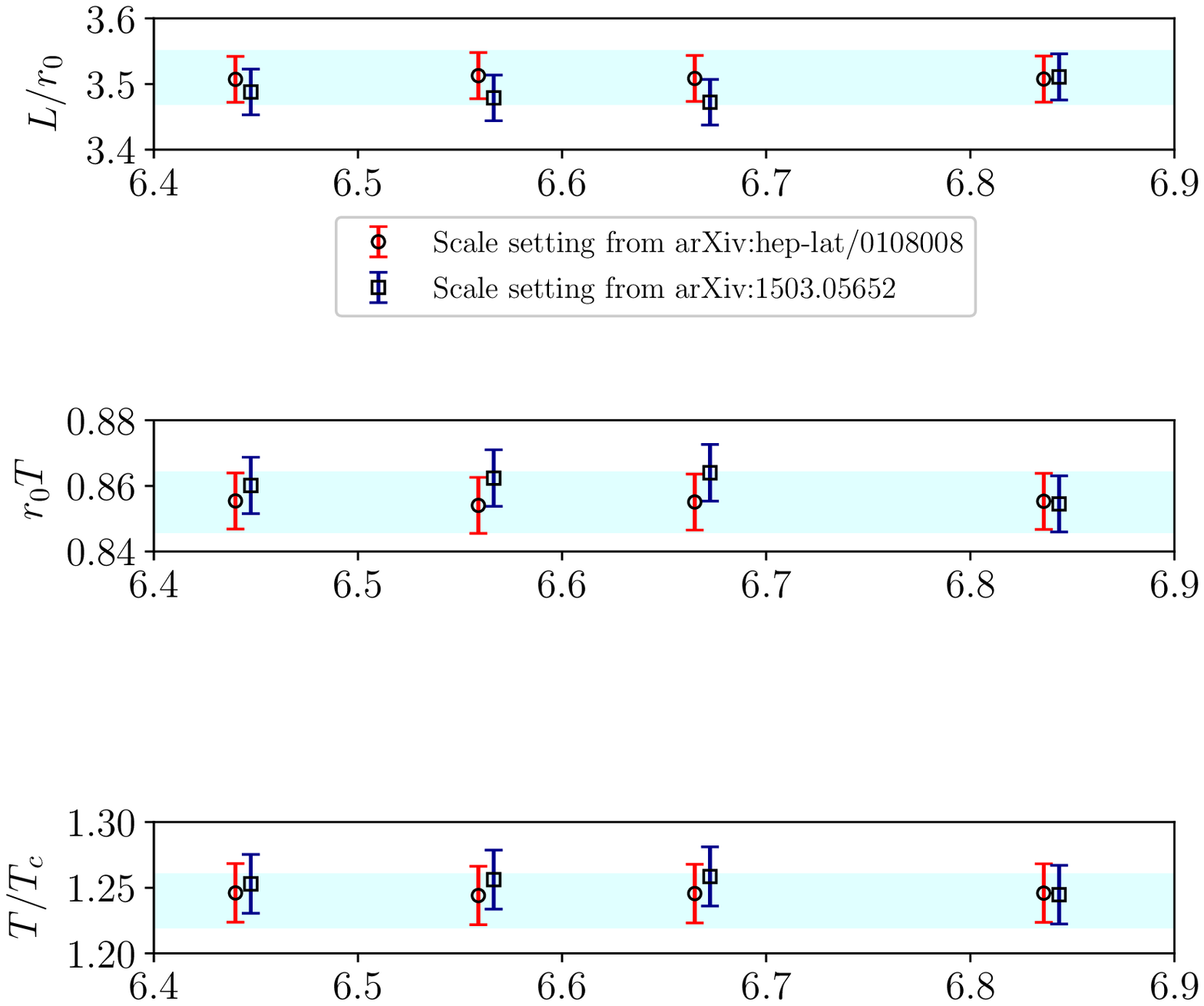}
\caption{Comparison of two different scale setting procedures, performed according to results of Ref.~\cite{Necco:2001xg} (round points) and Ref.~\cite{Francis:2015lha} (square points). The values along the LCP of $L/r_0 = 3.51(4)$, $r_0 T = 0.855(9)$ and $T/T_c = 1.24(2)$, determined from the scale setting used in this work, are represented as shaded areas.}
\label{fig:LCP_comp}
\end{figure}

The results of these two different scale setting procedure are summarized in Fig.~\ref{fig:LCP_comp}, where we compare the results for $a(\beta)/r_0$, $L/r_0$, $r_0 T$ and $T/T_c$. Note that, for both scale setting procedures, $r_0 T$ was converted into $T/T_c$ using the same critical temperature $T_c \simeq 287~\mathrm{MeV}$. As it can be appreciated, any observed deviation is smaller than the $\sim 1\%$ error, i.e., it stays within the quoted precision for the lattice spacing.

\section{Choice of the smearing width of the target function}\label{app:rate_vs_sigma}

As discussed in Sec.~\ref{sec:backus-gilbert}, the choice of the target function is a fundamental ingredient to assess the quality of the reconstruction of the spectral density via the modified Backus--Gilbert inversion method we applied in this work.

On general theoretical grounds, it can be shown, for sufficiently small smearing widths, that the dependence of the reconstructed quantity on $\sigma$ for an even target function in $\omega/\sigma$ can be expanded in powers of $\sigma^2$~\cite{Bulava:2021fre,Frezzotti:2023nun,Evangelista:2023vtl}:
\beq\label{eq:zerosigma_extr}
\Gamma_\sphal(\sigma) = \Gamma_\sphal(0) + \tilde{k} \, \sigma^2 + O(\sigma^4).
\eeq

\begin{figure}[!t]
\centering
\includegraphics[scale=0.5]{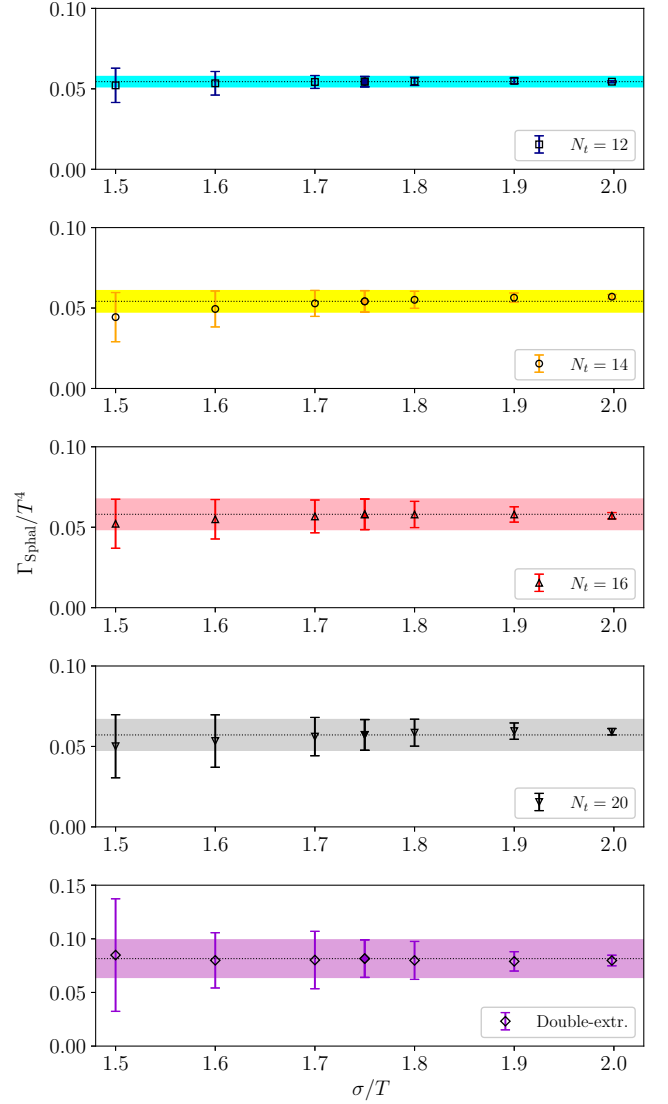}
\caption{Dependence of the sphaleron rate on the smearing width in physical units $\sigma/T$ in the range $1.5 \le \sigma/T < 2$ for all available lattice spacings and for approximately the same value of the smoothing radius $r_s T \propto n_\cool/N_t^2 \simeq 0.035$. In the last panel from the top we also show the determination of the rate from the double-extrapolated correlator as a function of the smearing width. Shaded areas represent the value for $\sigma/T=1.75$, which is the smearing width adopted in this work.}
\label{fig:rate_vs_sigma}
\end{figure}

In Fig.~\ref{fig:rate_vs_sigma} (top panels), we show how the rate depends on the choice of the smearing width $\sigma$ for all available lattice spacings, and for a single value of $n_\cool/N_t^2 \simeq 0.035$, which was chosen so as to stay well within the region where we observe a plateau in the continuum limit of $\Gamma_\sphal$ as a function of the smoothing radius, cf.~Fig.~\ref{fig:rate_zerocool}. In the bottom panel of Fig.~\ref{fig:rate_vs_sigma} we also show the $\sigma$-dependence of the spahleron rate obtaind from the inversion of the double-extrapolated correlator in the same ranges of smearing widths.

\begin{figure}[!t]
\centering
\includegraphics[scale=0.43]{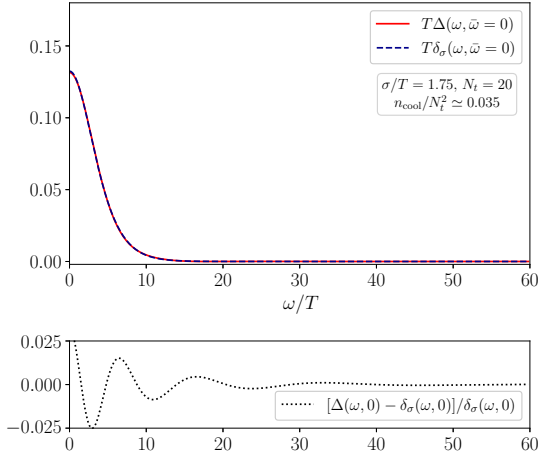}
\caption{Top panel: obtained resolution function $\Delta(\omega,\bar{\omega}=0)$ for $N_t=20$, $n_\cool/N_t^2\simeq 0.035$, compared with the target function $\delta_\sigma(\omega, \bar{\omega}=0)$ in Eq.~\eqref{eq:target}, with $\sigma/T=1.75$. Bottom panel: relative difference between resolution and target functions.}
\label{fig:resolution_function}
\end{figure}

As expected, as the target function gets more peaked, the errors increase, since the reconstruction becomes noisier. On the other hand, increasing the width of the target function diminishes the errors, as the spectral density is smeared over a larger region. Although in principle choosing too large values of $\sigma$ could potentially introduce undesired systematic effects in the sphaleron rate, we do not observe any sizeable systematic effect on our results for the sphaleron rate when varying the width of the target function in the range $1.5 \le \sigma/T < 2$. These findings are compatible with the theoretical expectation that, for an even target function like ours, cf.~Eq.~\eqref{eq:target}, linear corrections in $\sigma$ exactly vanish. Similar behaviors have also been observed in other works adopting the method of~\cite{Hansen:2019idp} to obtain other reconstructed quantities~\cite{Bulava:2021fre,Frezzotti:2023nun,Evangelista:2023vtl}.\\
\hspace*{0.4cm}In conclusion, thus, being our determinations of $\Gamma_\sphal/T^4$ essentially independent of $\sigma/T$ within the explored range, we took the results obtained for $\sigma/T=1.75$ (full points in Fig.~\ref{fig:rate_zerocool}) as our final results for the sphaleron rate. As it can be seen from Fig.~\ref{fig:rate_zerocool}, such choice always yields a safe and conservative estimate of the error in all cases.\\
\hspace*{0.4cm}Finally, in Fig.~\ref{fig:resolution_function}, we show the obtained resolution function $\Delta(\omega,\bar{\omega}=0)$ for $N_t=20$ and for the same smoothing radius used in Fig.~\ref{fig:rate_zerocool}, and we compare it with the chosen target function $\delta_\sigma(\omega, \bar{\omega}=0)$ in Eq.~\eqref{eq:target}, with $\sigma/T=1.75$. As it can be appreciated, the obtained reconstruction is excellent, as any relative difference between the obtained resolution function and the target one is in this case at most of $\sim2.5\%$.

\FloatBarrier

\end{document}